\documentclass[aps,pre,twocolumn,nofootinbib,tightenlines,superscriptaddress,nopacs,amsmath,amssymb,final,letterpaper]{revtex4-2}
\usepackage[utf8]{inputenc}
\usepackage{calc}
\usepackage{graphicx}
\usepackage{amsmath,amssymb,amsthm}
\usepackage{bbold}
\usepackage{bm}
\usepackage{color}
\usepackage[dvipsnames]{xcolor}
\usepackage{enumitem}
\usepackage{tikz}
\usepackage{float}
\usepackage{array}
\usepackage{cellspace}
\usepackage{booktabs}
\usepackage[colorlinks=true,linkcolor=blue,urlcolor=blue,citecolor=blue,anchorcolor=blue]{hyperref}
\usepackage{booktabs}
\usepackage{cellspace, hhline}

\DeclareMathOperator*{\argmin}{arg\,min}
\newcommand{\abs}[1]{\vert #1\vert}
\def\multiset#1#2{\ensuremath{\left(\kern-.3em\left(\genfrac{}{}{0pt}{}{#1}{#2}\right)\kern-.3em\right)}}
\renewcommand{\thetable}{\arabic{table}} 

\usepackage{makecell} 
\usepackage{ragged2e} 

\begin{document}
\setcitestyle{round} 

\title{Information theory for hypergraph similarity}

\author{Helcio \surname{Felippe}${}^{\ddag}$}
\affiliation{Department of Network and Data Science, Central European University, Vienna}

\author{Alec \surname{Kirkley}${}^{\ddag}$}
\email{alec.w.kirkley@gmail.com}
\affiliation{Institute of Data Science, University of Hong Kong, Hong Kong}
\affiliation{Department of Urban Planning and Design, University of Hong Kong, Hong Kong}
\affiliation{Urban Systems Institute, University of Hong Kong, Hong Kong}

\author{Federico \surname{Battiston}}
\email{battistonf@ceu.edu}
\affiliation{Department of Network and Data Science, Central European University, Vienna}
\affiliation{Department of AI, Data and Decision Sciences, Luiss University of Rome, Rome, Italy}

\def\thefootnote{\ddag}\footnotetext{These authors contributed equally to this work}

\date{\today}

\begin{abstract}
Comparing networks is essential for a number of downstream tasks, from
clustering to anomaly detection. Despite higher-order interactions
being critical for understanding the dynamics of complex systems,
traditional approaches for network comparison are limited to pairwise
interactions only.  Here we construct a general information theoretic
framework for hypergraph similarity, capturing meaningful
correspondence among higher-order interactions while correcting for
spurious correlations. Our method operationalizes any notion of
structural overlap among hypergraphs as a principled normalized mutual
information measure, allowing us to derive a hierarchy of increasingly
granular formulations of similarity among hypergraphs within and
across orders of interactions, and at multiple scales. We validate
these measures through extensive experiments on synthetic hypergraphs
and apply the framework to reveal meaningful patterns in a variety of
empirical higher-order networks. Our work provides foundational tools
for the principled comparison of higher-order networks, shedding light
on the structural organization of networked systems with
non-dyadic interactions.
\end{abstract}

\maketitle

\section{Introduction}

Comparing networked systems is central to a variety of downstream
tasks in the analysis of complex systems, with applications including
clustering, classification, and
regression~\cite{de2015structural,ok2020graph,attar2017classification}.
As a result, substantial research has been devoted to developing
measures that are capable of capturing similarities in salient
structural features of
networks~\cite{wills2020metrics,kriege2020survey}, with graph
similarity measures applied widely across scientific domains spanning
biology~\cite{sharan2006modeling},
chemistry~\cite{nikolova2003approaches},
neuroscience~\cite{mheich2020brain}, and
sociology~\cite{faust2002comparing} among others. 

\begin{figure*}[ht!]
\includegraphics[width=1.0\textwidth]{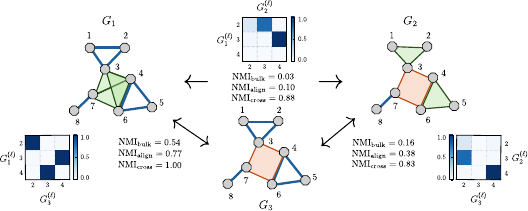}
\caption{
    \textbf{Hierarchy of information-theoretic measures for hypergraph similarity.}
    Hypergraphs $G_1$, $G_2$, and $G_3$ are defined on the same set of
    $N=8$ labeled nodes, with hypergraph layers $G_i^{(\ell)}$ indexed
    by $\ell\in\{2,3,4\}$ and illustrated as thick blue lines (dyads),
    green triangles (triplets), and orange squares (quadruplets).
    Heatmaps show the order-order mutual information between pairwise
    projections of hypergraph layers $G_i^{(\ell)}$ and $G_j^{(k)}$,
    for all $\ell,k\in \{2,3,4\}$. The three proposed hypergraph
    mutual information measures---$\text{NMI}_{\rm bulk}$,
    $\text{NMI}_{\rm align}$, $\text{NMI}_{\rm cross}$, which are
    derived using the general framework discussed in
    Sec.~\ref{sec:hierarchy}---are shown for each pair of
    hypergraphs. These measures assess the structural similarity
    between a pair of hypergraphs with increasingly detailed
    encodings to highlight structural overlaps at and across
    different hyperedge orders.
    }
\label{fig:fig1}
\end{figure*}

The majority of existing network similarity measures are tailored for
analyzing graphs consisting solely of pairwise interactions among the
entities comprising the nodes in the graph. However, it has been shown
through a vast body of recent work that pairwise interactions alone
are not sufficient for understanding the structure and dynamics
present in many real complex systems, wherein interactions often
involve groups of more than two
nodes~\cite{lambiotte2019networks,battiston2020networks,battiston2021physics,bianconi2021higher,bick2023higher}.
Hypergraphs, which generalize graphs to sets of edges containing any
number of nodes~\cite{berge1984hypergraphs}, provide a highly flexible
representation for modeling complex systems, allowing for more precise
modeling of collective phenomena in contagion and diffusion
processes~\cite{iacopini2019simplicial,neuhauser2020multibody,ferraz2024contagion,di2024dynamical},
synchronization and evolutionary
dynamics~\cite{millan2020explosive,zhang2023higher,civilini2024explosive},
and more.  

Despite the growing methodological toolkit for analyzing hypergraph
data~\cite{chodrow2021generative, veldt2023combinatorial,
ruggeri2023community}, there are relatively few measures for comparing
them~\cite{surana2022hypergraph,martino2020hyper,agostinelli2025higher}.
Many of these methods, including those based on vector
embeddings~\cite{bai2021hypergraph} and combinations of structural
features~\cite{feng2024hyper}, require the specification of free
parameters to which the results are highly sensitive, making them
challenging to apply in practice without substantial fine-tuning.
Meanwhile, methods based on spectral
properties~\cite{saito2022hypergraph}, path
lengths~\cite{agostinelli2025higher}, random
walks~\cite{bai2012jensen}, and
graphlets~\cite{lugo2021classification} are capable of capturing
complex structural dependencies without tunable parameters, but impose
a computational complexity that is at least quadratic in the number of
nodes in the network, causing them to scale poorly to large systems.
Additionally, as many of these methods incorporate ad hoc structural
features into the similarity calculation with no clear fundamental
principles motivating the modeling choices, they provide results that
are hard to interpret and may not generalize well to hypergraphs
across application domains without substantial modification.  

By focusing on the connection between structural regularities in data
and its compressibility~\cite{mackay2003information}, information
theory provides a principled foundation on which to build methods for
extracting salient structural features in network data in a
nonparametric manner. In particular, the minimum description length
(MDL) principle, which states that the best model for a dataset is the
one that allows for its shortest description in terms of bits of
information~\cite{rissanen1978}, is at the heart of many unsupervised
methods for understanding large-scale network structure, including
methods for
clustering~\cite{peixoto2019bayesian,kirkley2022spatial,morel2024bayesian},
reconstruction and
denoising~\cite{peixoto2018reconstructing,kirkley2023compressing}, and
identifying influential or highly connected groups of
nodes~\cite{gallagher2021clarified,kirkley2024identifying}.  By aiming
to compress network data based on information encodings that exploit
certain structural regularities of interest---e.g., communities,
highly connected hub nodes, etc---MDL-based methods can extract
statistically significant structure in graphs while ignoring spurious
regularities that arise from statistical noise. 

Given its explanatory power and flexibility, information theory has
been used in a number of existing works to construct measures for
graph
comparison~\cite{de2015structural,de2016spectral,escolano2017mutual,corso2020mutual},
some of which have explicitly utilized the MDL
principle~\cite{coupette2021graph,felippe2024network}. Formulating the
problem of graph comparison using the MDL principle allows for fully
nonparametric methods that are principled and interpretable.  The MDL
principle thus provides an ideal framework with which to develop
measures of hypergraph similarity.

Here we \textcolor{black}{introduce} a framework for constructing
principled and interpretable information-theoretic hypergraph
similarity measures utilizing the MDL principle. 
\textcolor{black}{Within this framework, we}
derive a series of similarity measures capturing the multiscalar,
nested nature of higher-order interactions in an increasingly granular
manner. We extend these measures to compute the similarity between a
pair of hypergraphs under an arbitrary coarse-graining of the nodes,
permitting comparison of interaction patterns across hypergraphs at a
desired scale of interest while ignoring fluctuations below the
specified scale. Through a range of experiments on real and synthetic
hypergraphs, we demonstrate that our framework allows for measures
that capture nuanced aspects of structural similarity at multiple
scales while remaining robust to statistical noise.

\section{Measures}

Analogous to the construction of information theoretic measures for
comparing partitions \cite{newman2020improved,jerdee2024mutual}, one
can construct entropy and conditional entropy measures for any pair of
discrete objects by considering different encodings of their structure
in an information transmission process. The most natural way to do
this is to assign multiple fixed-length codes to transmit the
information in the objects at increasing resolution, starting with
coarse summary statistics and ending with the final transmission of
their detailed structure. The information shared by these objects in
their structural overlap can then be quantified using mutual
information, which is the amount of information saved in specifying
one object when the other is known. 

Here we develop a general framework for hypergraph similarity measures
that take the form of normalized mutual information scores that
quantify the amount of shared information between a pair of
hypergraphs. By specifying different encoding schemes---that is,
different methods for transmitting the network data---we derive a
family of mutual information-based hypergraph similarity measures that
capture both intra- and cross-order similarity among hypergraphs. We
extend these measures to capture similarity at higher scales using
arbitrary coarse-grainings of the nodes.

Given that hypergraphs have higher-order interactions than the simple
pairwise interactions in normal graphs, there are a number of
adaptations one can make to highlight different aspects of similarity
within and across different interaction orders. Here we explore these
generalizations in increasing order of complexity to provide a
hierarchy of hypergraph mutual information measures suitable for the
comparison of empirical higher-order data.

\textcolor{black}{
\subsection{Mutual information for graph similarity}\label{sec:MIgraph}
} 

Here we discuss how to construct a simple mutual information measure
for graph comparison \cite{felippe2024network}, which will help to
motivate the construction of our hypergraph mutual information
measures.  For simplicity, we consider input graphs~$G_1$,~$G_2$
that are unweighted and simple (undirected and without self- or
multi-edges), although in principle it is straightforward to
extend the measure to these cases as well by accounting for
directed and self-edges by allowing additional valid edge
positions or multi-edges with multiset combinatorics (as discussed
in~\cite{felippe2024network}). The graphs~$G_i$ with
$i\in\{1,2\}$ exist on the same set of $N$ labeled nodes, allowing
their structural overlap to be unambiguously computed in the
computation of the mutual information, and can be represented as
sets of $E_1$ and $E_2$ (sorted) tuples respectively in an
edgelist representation.

Similar to node partitions, one can construct entropy and conditional
entropy measures for the entire graphs $G_1$, $G_2$ that consider
different encodings of their information for transmission of these
objects to a receiver~\cite{newman2020improved,jerdee2024mutual}.  The
entropy (or description length) of the graph~$G_i$ under a particular
encoding scheme is the amount of information it takes to specify $G_i$
after minimizing the codelength over any intermediate representations
of $G_i$---e.g. node communities---for optimal compression. By the
Kraft inequality~\cite{cover2012elements}, for any properly normalized
probability distribution~$P(G)$ over some set of valid
graphs~$\mathcal{G}$, there exists a uniquely decodable prefix code
over bitstrings with codelength~$-\log_2P(G)$ for the graph~$G$.
Therefore, to assign a codelength to $G_i$ to compute its entropy, all
we need is a probability distribution over the possible
configurations~$\mathcal{G}$ that $G_i$ may take. 

The simplest---and most widely used---encoding is the fixed-length
code, which in this case assigns all graphs~$G\in \mathcal{G}$ the
same codelength~$\abs{\mathcal{G}}^{-1}$~\cite{mackay2003information}.
To highlight some structural property of interest about a graph, such
as its community structure, one can construct a sequence of
fixed-length codes---corresponding to a hierarchical uniform Bayesian
prior~\cite{peixoto2019bayesian}---that utilize this structural
property to achieve compression by reducing the set of possible graphs
$\mathcal{G}$ in the final encoding step as much as possible. At the
same time, one aims to not overcomplicate the encoding to avoid
wasting information describing each step. The optimal balance,
according to the MDL principle, is achieved when the total codelength
(description length) of the observed graph and intermediate encoding
steps is minimized.

We can first assume that the receiver knows the total number of
nodes~$N$ and number of edges~$E_1$, $E_2$ in the two graphs. (These
quantities require comparatively negligible information content to
specify, so we can safely ignore them.) The simplest encoding is then
the fixed-length code over all graphs compatible with these known
constraints. There are ${{N\choose 2}\choose E_i}$ possible simple
graphs of $E_i$ edges on $N$ labeled nodes, and so the entropy
(codelength) of graph~$G_i$ under this encoding is just
\begin{align}\label{graph-ent}
    \text{H}_{\text{graph}}(G_i) = \log {{N\choose 2}\choose E_i},    
\end{align}
where we have abbreviated $\log \equiv \log_2$ for brevity.  

In a similar manner to Eq.~(\ref{graph-ent}), we can construct a
conditional entropy measure between $G_1$ and $G_2$, which tells us
the amount of information to describe $G_2$ given that $G_1$ is known
by the receiver (or vice versa). In this case, the Kraft inequality
tells us that any probability distribution~$P(G\vert G_1)$ over
graphs~$G$ given the known graph~$G_1$ corresponds to a valid encoding
for $G_2$. However, to achieve compression of $G_2$ when $G_1$ is
known, we must specify some measure of overlap among the two
graphs---without this overlap, our knowledge of $G_1$ is
uninformative and thus not useful for compression. When specifying
the overlap we also have considerable modeling freedom to
highlight any structural features of interest. In the simplest
case, we can use the set~overlap among $G_1$ and $G_2$, which
counts the number of edges they have in common. Denoting this
overlap as
\begin{align}\label{overlap}
    E_{12} = \abs{G_1 \cap G_2},    
\end{align}
we have that there are ${E_1\choose E_{12}}$ possible configurations
of the $E_{12}$ overlapping edges given that they must be a subset of
the $E_1$ edges in the known $G_1$. Specifying the overlap~$\abs{G_1
\cap G_2}$ therefore costs us $\log {E_1\choose E_{12}}$ bits. After
receiving this overlap set, the receiver can exclude the $E_1-E_{12}$
edges in $G_1$ that are not included in the overlap from the
possibility of occurring in $G_2$. Thus, there are ${N\choose 2}-E_1$
remaining edges that could occur in $G_2$, of which $E_2-E_{12}$ are
present, excluding the $E_{12}$ we already know from the overlap.
Therefore, specifying $G_2$ given the known overlap~$\abs{G_1 \cap
G_2}$ requires $\log {{N\choose 2}-E_1 \choose E_2-E_{12}}$~bits of
information. Putting it all together, we have that the conditional
entropy of $G_2$ given $G_1$ is
\begin{align}\label{graph-CE}
    \text{H}_{\text{graph}}(G_2\vert G_1) = \log {E_1\choose E_{12}}{{N\choose 2}-E_1 \choose E_2-E_{12}}.    
\end{align}
Note that, as with the entropy of Eq.~(\ref{graph-ent}), the
conditional entropy of Eq.~(\ref{graph-CE}) depends on the encoding
one chooses---in other words, the way to measure the overlap among
$G_2$ and $G_1$. We will see in Sec.~\ref{sec:hierarchy} that this
allows us to capture similarity among hypergraphs within and across
different orders of hyperedges, as well as at different scales of
interest.

The difference between Eq.~(\ref{graph-ent}) and Eq.~(\ref{graph-CE})
quantifies the amount of information we save about $G_2$ by first
knowing $G_1$ and its overlap with $G_2$. This is called the
\emph{mutual information} (MI) of $G_1$ and
$G_2$~\cite{felippe2024network}, thus
\begin{align}\label{graph-MI}
    \text{MI}_{\text{graph}}(G_1;G_2) = \text{H}_{\text{graph}}(G_2)-\text{H}_{\text{graph}}(G_2\vert G_1),   
\end{align}
and can be used directly as a measure of similarity among the two
graphs. When $G_1$ and $G_2$ are very similar---i.e., have a high
overlap $E_{12}$---Eq.~(\ref{graph-MI}) will also be high, since
knowing $G_1$ and the overlap will substantially constrain the number
of possibilities for $G_2$ (the conditional entropy $\text{H}(G_2\vert
G_1)$ will be low). On the other hand, when $G_1$ and $G_2$ are very
different (have low overlap~$E_{12}$ and therefore high
$\text{H}(G_2\vert G_1)$), the mutual information will be low because
we do not save much information about $G_2$ by knowing $G_1$ and the
overlap.  

Through the Vandermonde identity~\cite{roberts2024applied}, we have
\begin{align}\label{vandermonde}
    \log {\sum_n y_n \choose \sum_n x_n} \geq \sum_n \log {y_n\choose x_n}
\end{align}
for any sequences~$\{x_n\},\{y_n\}$ of non-negative integers with
$y_n\geq x_n$, which implies that
$\text{MI}_{\text{graph}}(G_1;G_2) \geq 0$, such that we will
always save information about $G_2$ by specifying $G_1$ and the
overlap first. (This is a combinatorial version of the concept
that conditioning always reduces
entropy~\cite{mackay2003information}.) Equation~(\ref{graph-MI})
also has the nice property of being symmetric in the
graphs~$G_1,G_2$, as one can show that
\mbox{$\text{H}(G_2)-\text{H}(G_2\vert
G_1)=\text{H}(G_1)-\text{H}(G_1\vert G_2)$}. These two
properties---non-negativity and symmetry---are often desirable for
mutual information measures, but are not strictly necessary. For
example, if one wants to account for the information required to
transmit the overlap itself to provide a more accurate accounting
of the conditional entropy and reduce finite-size biases, it may
sacrifice the symmetry and non-negativity of the MI depending on
how the overlap is
encoded~\cite{newman2020improved,jerdee2024mutual,jerdee2025normalized,kirkley2025transfer}.
In~\cite{felippe2024network}, as well as this paper, we ignore the
information content of specifying the overlap to ensure
non-negativity of the mutual information measures. However, for
the hypergraph case, we will find that breaking the symmetry
allows for more encoding flexibility to capture different aspects
of overlap.\\

\subsection{Hypergraph normalized mutual information framework}
\label{sec:hypergraph-NMI}

We consider input hypergraphs~$G_1$, $G_2$ on the same set of $N$
(aligned) nodes that are unweighted and simple, i.e. have no multi- or
self-edges. Extensions of our measures to relax the unweighted and
simple hypergraph assumptions are conceptually straightforward but
involve more complex multiset combinatorics (see
SM~\ref{app:multiscale}). In the hypergraph case, $G_1$ and $G_2$ can
be represented as edge sets of tuples with two or more nodes, ordered
by node index to impose undirectedness. Since hyperedges of different
orders have qualitatively different interpretations and impacts on
network dynamics~\cite{battiston2020networks}, each hypergraph~$G_i$
can be decomposed into ``layers''~$\mathcal{L}=\{2,...,L\}$ such that
the layer~$G_i^{(\ell)}$ contains all hyperedges of
size~(order)~$\ell$ in $G_i$, and $L$ is the maximum order of
hyperedges across the two hypergraphs~$G_1$, $G_2$. If no hyperedge of
size~$\ell$ exists in $G_i$, we set $G^{(\ell)}_i=\{~\}$.
\textcolor{black}{In Sec.~\ref{section:multiplex} of the SM we discuss
the case in which the hypergraphs have aligned node labels but
different node sets, and which preprocessing options are available
prior to computing their similarity.}

To account for the nestedness of interactions in our measures, a
feature observed in many empirical systems with higher-order
interactions~\cite{lotito2022higher,larock2023encapsulation,landry2024simpliciality,gallo2024higher},
we can also let each layer~$G_i^{(\ell')}$ be ``projected'' down onto
hyperedges of size~$\ell \leq \ell'$ by taking the set of all unique
sub-tuples of size~$\ell$ within the tuples of $G_i^{(\ell')}$. We
denote the projection from layer~$\ell'$ to $\ell$ as $G^{(\ell'\to
\ell)}_i$, with the convention~$G^{(\ell\to \ell)}_i=G_i^{(\ell)}$.
For example, if \mbox{$G_i=\{(0,1,2),(1,2)\}$}, we would have
$G^{(3)}_i=\{(0,1,2)\}$, $G^{(2)}_i=\{(1,2)\}$, and $G^{(3\to
2)}_i=\{(0,1),(1,2),(0,2)\}$. We will let the size of (number of
hyperedges in) any set~$G^{(x)}_i$ be denoted with
$E^{(x)}_i=\abs{G^{(x)}_i}$, such that in the previous example we have
$E^{(3)}_i=1$, $E^{(2)}_i=1$, $E^{(3\to 2)}_i=3$, and
$E_i=\sum_{\ell\in \mathcal{L}}E_i^{(\ell)}=2$.  

We can now define hypergraph mutual information measures by
considering the transmission of the hypergraph~$G_2$ by itself as well
as given the known~$G_1$ and some measure(s) of overlap between the
two hypergraphs. For generality, we can consider
\begin{align}\label{general-ent}
    \text{H}_{c}(G_i)
    = \log\big[\text{\# possible $G_i$ under encoding $c$}\big],
\end{align}
and 
\begin{align}\label{general-CE}
    \text{H}_{c}(G_j\vert G_i)
    = \log\big[\text{\# possible $G_j$ under $c$ given $G_i$}\big].
\end{align}
These expressions reflect the fact that the number of possible
configurations of a hypergraph, and hence its entropy/conditional
entropy, depend on what encoding scheme we use---in particular, which
constraints are assumed to be known by the receiver under the encoding
scheme, and how the encoding scheme defines the overlap among $G_1$
and $G_2$. For example, if we let $c=\text{``graph''}$ be the encoding
described in Sec.~\ref{sec:MIgraph}, we recover Eq.~(\ref{graph-ent})
from the entropy in Eq.~(\ref{general-ent}), and Eq.~(\ref{graph-CE})
from the entropy in Eq.~(\ref{general-CE}). Given
Eqs.~(\ref{general-ent}) and (\ref{general-CE}), we can then construct
a mutual information measure between $G_1$ and $G_2$, thus
\begin{align}\label{general-MI}
    \text{MI}_{c}(G_1;G_2) &= \text{H}_{c}(G_2) - \text{H}_{c}(G_2\vert G_1). 
\end{align}

To have a uniform scale on which to compare hypergraphs, it is useful
to normalize Eq.~(\ref{general-MI}) so that it falls in the
range~$[0,1]$, equaling $1$ when $G_1$ and $G_2$ are identical and a
value near $0$ when $G_1$ and $G_2$ are completely different from each
other (i.e. have little overlap). Examining Eq.~(\ref{general-MI}), we
can immediately see that \mbox{$0\leq \text{MI}_{c}(G_1;G_2)\leq
\text{H}_{c}(G_2)$}.  The upper bound on the MI results from the fact
that the number of configurations of $G_2$ \emph{without} any
additional constraints from $G_1$ ($2^{\text{H}_{c}(G_2)}$) must be at
least as large as the number of configurations of $G_2$ \emph{with}
additional constraints from $G_1$ ($2^{\text{H}_{c}(G_2\vert G_1)}$).
And the lower bound on the MI results from the non-negativity of the
conditional entropy, since its argument (a positive count value) is
always at least equal to $1$.  To allow for full generality in the
encodings~$c$, we will allow $\text{MI}_c$ to potentially be
asymmetric, in which case we can construct a symmetric normalized MI
measure by taking the maximum of the fractional shared information
when considering transmitting $G_2$ from $G_1$ and $G_1$ from $G_2$.
This gives a normalized mutual information measure (NMI) of
\begin{align}\label{general-NMI}
    \text{NMI}_{c}(G_1,G_2) &= \text{max}\left\{\frac{\text{MI}_{c}(G_1;G_2)}{\text{H}_{c}(G_2)},\frac{\text{MI}_{c}(G_2;G_1)}{\text{H}_{c}(G_1)}\right\}\\
    &=1-\text{min}\left\{\frac{\text{H}_{c}(G_2\vert G_1)}{\text{H}_{c}(G_2)},\frac{\text{H}_{c}(G_1\vert G_2)}{\text{H}_{c}(G_1)}\right\}. 
\end{align}
The NMI measure in Eq.~(\ref{general-NMI}) is highly flexible,
providing a general framework for constructing hypergraph similarity
measures.

Equation~(\ref{general-NMI}) provides a natural mechanism for
assessing the similarity among hypergraphs~$G_1$, $G_2$ in a manner
that is robust to statistical noise. Real-world hypergraphs are
typically extremely sparse, only containing a vanishing fraction of
the ${N\choose \ell}$ possible hyperedges at each order~$\ell$, with
the sparsity becoming more pronounced as we increase
$\ell$~\cite{battiston2020networks}. Thus, two hypergraphs~$G_1$,
$G_2$ that are completely uncorrelated---e.g., generated as
independent random hypergraphs on $E_1$ and $E_2$ hyperedges
respectively---will have an overlap that approaches zero for large $N$
for any overlap measure that is based on the number of shared tuples
among the two hypergraphs or their individual layers (e.g.
Eq.~(\ref{overlap})). We therefore have that $G_i$ and the overlap
place very weak constraints on $G_j$, so that
\mbox{$\text{H}_{c}(G_j)\approx \text{H}_{c}(G_j\vert G_i)$} and
\mbox{$\text{NMI}_c\approx 0$}. We will more concretely see how
this manifests itself in the experiments in
Sec.~\ref{sec:results}. 

Table~\ref{tab:framework} summarizes the structure of the general NMI
framework we propose for constructing hypergraph similarity measures
from fundamental information theoretic principles. While we explore
three specific encodings for a natural hierarchy of similarity
measures in this paper, our framework applies much more broadly to any
meaningful encoding of hypergraph structure.  

\renewcommand{\arraystretch}{1.4}
\setlength{\tabcolsep}{4pt}
\begin{table}[h]
\centering
\begin{tabular}{|c|c|}
\hline
\textbf{Measure} & \textbf{Description} \\ \hline
$\text{H}_{c}(G_i)$ &
\makecell[l]{
            \begin{minipage}[l]{0.32\textwidth}
            \justifying\vspace{0.125cm}
            Entropy of hypergraph~$G_i$. Amount of information to transmit $G_i$ using an arbitrary lossless encoding~$c$.
            \vspace{0.125cm}
            \end{minipage}
            } \\ \hline
$\text{H}_{c}(G_j\vert G_i)$ &
\makecell[l]{
            \begin{minipage}[l]{0.32\textwidth}
            \justifying\vspace{0.125cm}
            Conditional entropy of hypergraph $G_j$ given hypergraph $G_i$.
            Amount of information to transmit $G_j$ using lossless encoding~$c$
            when receiver has knowledge of both $G_i$ and a measure of its
            overlap with~$G_j$.
            \vspace{0.125cm}
            \end{minipage}
            } \\ \hline
$\text{MI}_{c}(G_i;G_j)$ &
\makecell[l]{
            \begin{minipage}[l]{0.32\textwidth}
            \justifying\vspace{0.125cm}
            Mutual information of ~$G_i$ and~$G_j$ (Eq.~(\ref{general-MI})). Amount of
            information saved when transmitting $G_j$ after knowing $G_i$, under
            encoding~$c$. Can be asymmetric in general.
            \vspace{0.125cm}
            \end{minipage}
            } \\ \hline
$\text{NMI}_{c}(G_1,G_2)$ &
\makecell[l]{
            \begin{minipage}[l]{0.32\textwidth}
            \justifying\vspace{0.125cm}
            Normalized mutual information of $G_1$ and~$G_2$
            (Eq.~(\ref{general-NMI})). Fraction of information saved when
            transmitting one hypergraph from another using encoding~$c$, under
            the more efficient order of transmission. Manifestly symmetric
            hypergraph similarity measure bounded in $[0,1]$.
            \vspace{0.125cm}
            \end{minipage}
            } \\ \hline
\multicolumn{2}{c}{} \\[-1ex] 
\hline
\textbf{Encoding, $c$} & \textbf{Description} \\ \hline
bulk &
\makecell[l]{
            \begin{minipage}[l]{0.32\textwidth}
            \justifying\vspace{0.125cm}
            Transmits hypergraphs by specifying hyperedges of all sizes at
            once. Only accounts for intra-order similarity and is only robust
            to statistical noise for homogeneous layer densities.
            \vspace{0.125cm}
            \end{minipage}
            } \\ \hline
align &
\makecell[l]{
            \begin{minipage}[l]{0.32\textwidth}
            \justifying\vspace{0.125cm}
            Transmits hypergraphs by specifying hyperedges of each layer
            separately, with layer $G^{(\ell)}_j$ transmitted using layer
            $G^{(\ell)}_i$ (and vice versa). Only accounts for intra-order
            similarity but is robust to statistical noise for any layer
            densities.
            \vspace{0.125cm}
            \end{minipage}
            } \\ \hline
cross &
\makecell[l]{
            \begin{minipage}[l]{0.32\textwidth}
            \justifying\vspace{0.125cm}
            Transmits hypergraphs by specifying hyperedges of each layer
            separately, with layer $G^{(\ell)}_j$ transmitted using any layer
            $G^{(k)}_i$ for $k \geq \ell$. Accounts for both intra- and
            cross-order similarity and is robust to statistical noise for any
            layer densities.
            \vspace{0.0625cm}
            \end{minipage}
            } \\ \hline
\end{tabular}
\caption{
    Normalized mutual information framework for constructing
    hypergraph similarity measures, with descriptions of the encodings
    used in the proposed hierarchy of NMI measures in
    Sec.~\ref{sec:hierarchy}.
}
\label{tab:framework}
\end{table}

\subsection{A hierarchy of hypergraph similarity measures}
\label{sec:hierarchy}

Perhaps the simplest encoding $c$ one can consider is one in which all
hyperedges are transmitted at once. We will call this the ``bulk''
encoding to reflect the one-step transmission. In this case, following
the reasoning in Sec.~\ref{sec:MIgraph}, the entropy of each
hypergraph is 
\begin{align}\label{bulk-ent}
    \text{H}_{\text{bulk}}(G_i) = \log {2^N-N-1 \choose E_i}.    
\end{align}
Here, $2^N-N-1$ is the number of possible hyperedges of order at least
$2$ on $N$ nodes, of which we must choose $E_i$ hyperedges to fully
specify $G_i$. In this bulk transmission, the relevant overlap among
$G_1$ and $G_2$ which will be utilized for constructing a conditional
entropy is just
\begin{align}
    E_{12} = \abs{G_1\cap G_2},    
\end{align}
in which the entire sets~$G_1$ and $G_2$ are compared at the level of
their constituent tuples (analogous to Eq.~(\ref{overlap})). Then, the
conditional entropy under this bulk encoding scheme is given by the
logarithm of the number of ways $G_j$ may be configured given its
overlap $E_{12}$ with $G_i$, for $i,j\in \{1,2\}$. To transmit $G_j$
given knowledge of $G_i$, we must first specify which subset of
$E_{12}$ hyperedges among the $E_i$ hyperedges in $G_i$ form the
overlap among the two hypergraphs. Then we must specify the subset of
$E_j-E_{12}$ remaining hyperedges in $G_j$ that are found among the
$(2^N-N-1)-E_i$ possible hyperedges outside of $G_i$. This gives a
conditional entropy of 
\begin{align}\label{bulk-CE}
    \text{H}_{\text{bulk}}(G_j \vert G_i) = \log {E_i \choose E_{12}}{(2^N-N-1) - E_i \choose E_j-E_{12}}    
\end{align}
for $i,j\in \{1,2\}$. The NMI between the hypergraphs~$G_1$ and $G_2$
under this bulk encoding is then given by substituting
Eqs.~(\ref{bulk-ent})~and~(\ref{bulk-CE}) into
Eq.~(\ref{general-NMI}).

Although the bulk encoding provides perhaps the most intuitive way to
construct a hypergraph NMI measure using Eq.~(\ref{general-NMI}), it
has one critical limitation in practice: it considers \emph{any}
subset of the full space of $2^N-N-1$ possible hyperedges over $N$
nodes to be a valid hypergraph configuration when computing the
entropies. Since real hypergraphs often have increasing sparsity as we
go to higher and higher layers, the bulk encoding is very inefficient
for real hypergraphs~$G_i$ since it wastes a substantial amount of
space in its codebook assigning bitstrings to hypergraphs that we
are unlikely to ever observe. This issue manifests itself, in all
but the smallest hypergraphs, with an exaggerated level of
similarity between hypergraphs with very little overlap. In this
case, since $E_i/(2^N-N-1)$ is extremely small, $G_1$ and $G_2$
appear as if they share a substantial amount of information for
any non-zero overlap~$E_{12}>0$, as it is so unlikely that they
have any overlap given the enormous space of all possible
hypergraphs considered.

A simple way to correct for this issue is to consider the transmission
of each layer of the hypergraphs separately, which for the conditional
entropy requires an encoding that attributes similarity to the
hypergraphs at different layers separately as well. We will use the
notation~$c=\text{``align''}$ for the simplest variant of this
layer-wise encoding, which computes the similarity among $G_1$ and
$G_2$ using overlaps between layers of the same order only. There are
${N\choose \ell}$ possible hyperedges in layer~$\ell$, so if the
receiver knows there are $E^{(\ell)}_i$ hyperedges in layer~$\ell$, we
can transmit this layer using $\log {{N\choose \ell} \choose
E^{(\ell)}_i}$ bits. To transmit all layers separately, we thus need 
\begin{align}\label{align-ent}
    \text{H}_{\text{align}}(G_i) = \sum_{\ell\in \mathcal{L}}\log {{N\choose \ell} \choose E^{(\ell)}_i}    
\end{align}
bits. An analogous expression can be constructed for the conditional
entropy under this encoding by adjusting Eq.~(\ref{graph-CE}) for each
layer separately, thus
\begin{align}\label{align-CE}
    \text{H}_{\text{align}}(G_j\vert G_i) = \sum_{\ell\in \mathcal{L}}\log {E^{(\ell)}_i \choose E^{(\ell)}_{12}}{{N\choose \ell}-E^{(\ell)}_i \choose E^{(\ell)}_j-E^{(\ell)}_{12}},       
\end{align}
where 
\begin{align}
    E^{(\ell)}_{12} = \abs{G_1^{(\ell)}\cap G_2^{(\ell)}}    
\end{align}
is the overlap of the $\ell$-th layer in $G_1$ and the $\ell$-th layer
in $G_2$. Subbing into Eq.~(\ref{general-NMI}) then gives an NMI
measure $\text{NMI}_{\text{align}}(G_1,G_2)$ between hypergraphs $G_1$
and $G_2$ under this refined encoding. By assessing the similarity at
each layer separately to construct the conditional entropy, this
encoding also naturally accounts for sparsity differences across the
layers and provides a more granular description of similarity among
two hypergraphs than the bulk encoding. 

Both the ``bulk'' and ``align'' encodings are capable of capturing
similarity between hypergraphs \textcolor{black}{occurring}
\emph{within} the same order of hyperedges (i.e., \emph{intra-order}
similarity), but fail to capture similarity between hypergraphs that
can occur \emph{across} different orders of hyperedges (i.e.,
\emph{cross-order} similarity). For instance, dyadic interactions in
hypergraph $G_1$ might be similar to (subsets of) larger groups
interactions in $G_2$, but the two previous measures would assign to
$G_1$ and $G_2$ a low similarity score.
It is therefore useful to refine the encoding formulation to construct
a more flexible measure, able to capture higher-order similarity not
only within but also across multiple orders of interaction
simultaneously.

The unnormalized MI measures $\text{MI}_{\text{bulk}}(G_1;G_2)$
and $\text{MI}_{\text{align}}(G_1;G_2)$ (Eq.~(\ref{general-MI}))
will be symmetric in the input hypergraphs $G_1,G_2$. Thus, for
these cases the normalization of Eq.~(\ref{general-NMI}) is
equivalent to normalizing the MI by the smaller of the two
hypergraph entropies
$\text{min}(\text{H}_{c}(G_1),\text{H}_{c}(G_2))$. While this
symmetry is often desirable for an NMI measure of similarity,
allowing the MI to be asymmetric by using the more general form of
the NMI in Eq.~(\ref{general-NMI}) allows for greater flexibility
in the encodings one uses. Specifically, it allows for the
transmission of information across nested hyperedges of different
orders, since the information to transmit lower-order hyperedges
from higher-order hyperedges is different (typically much lower)
than the information to transmit higher-order hyperedges from
lower-order hyperedges---there are many higher-order hyperedges
compatible with (i.e. that are a superset of) a given set of
lower-order hyperedges, and higher-order interactions cannot be
uniquely determined from lower-order interactions
alone~\cite{larock2025exploring}. Meanwhile, there are
comparatively few lower-order hyperedges that are a subset of a
given higher-order hyperedge, reducing the information cost to go
``downward'' from higher- to lower-order relative to ``upward''
(the opposite direction). This natural asymmetry that arises in
cross-layer encodings is accommodated by the normalization of
Eq.~(\ref{general-NMI}), as it takes the better of the two
directions when quantifying similarity.

Beyond this inherent information asymmetry that typically favors
downward transmission in a cross-layer encoding, it is also
computationally much more efficient to consider only the downward
direction of transmission from a higher-order layer $k$ to a
lower-order layer $\ell$ and not vice versa. There are fast ways
to exactly compute the cardinality of a projection $G_i^{(k\to
\ell)}$ as well as its overlap with a lower order layer
$G_j^{(\ell)}$ of another hypergraph, see SM~\ref{app:algorithm}
for details. Meanwhile, there is likely no fast algorithm for the
reverse direction of transmission unless one sacrifices a
substantial amount of data compression. This is because an
efficient encoding will require using multiple hyperedges in
$G_j^{(\ell)}$ to transmit each hyperedge in
$G_i^{(k)}$---otherwise, there are at least ${N-\ell\choose
k-\ell}$ remaining choices for the nodes in each hyperedge of
$G_i^{(k)}$, since only $\ell$ nodes can be accounted for by a
single edge in $G_j^{(\ell)}$. The resulting encoding would
have a non-negligible information cost to map each
higher-order hyperedge in $G_i^{(k)}$ to a set of lower-order
hyperedges in $G_j^{(\ell)}$, and would also require a
matching algorithm for optimizing the encoding cost,
sacrificing the computational efficiency of the method.

Given the previous considerations of information asymmetry and
computational complexity, we develop a cross-layer encoding that
permits the transmission of layer~$G_j^{(\ell)}$ from any layer
$G_i^{(k)}$, so long as $\ell \leq k$.
To allow for a layer~$G_i^{(k)}$ in $G_i$ to aid in the transmission
of a layer~$G_j^{(\ell)}$ of a different order in $G_j$, we consider
the overlap of the projected layer~$G_i^{(k\to \ell)}$ and
$G_j^{(\ell)}$.  We can therefore define an overlap measure for
conditional entropies that incorporates cross-layer similarity as 
\begin{align}\label{overlap-cross}
    E^{(k\to \ell)}_{i\to j} = \abs{G_i^{(k\to \ell)}\cap G_j^{(\ell)}}, ~~k\geq \ell.
\end{align}
Modifying Eq.~(\ref{graph-CE}) appropriately then gives the layer-wise
conditional entropy
\begin{align}
    \log {E^{(k\to \ell)}_i \choose E^{(k\to \ell)}_{i\to j}}{{N\choose \ell} - E^{(k\to \ell)}_i \choose E^{(\ell)}_j-E^{(k\to \ell)}_{i\to j}}.
\end{align}
Now, if we aim to transmit $G_j$ in the most efficient way possible
under this encoding structure, we should transmit $G_j^{(\ell)}$ from
the layer $k_i^{(\ell)}$ in $G_i$ under which this layer-wise
conditional entropy is minimized. More formally, the best
layer~$k_i^{(\ell)}$ is given by
\begin{align}\label{best-layer}
    k_i^{(\ell)} = \argmin_{k\geq \ell}\left\{ \log {E^{(k\to \ell)}_i \choose E^{(k\to \ell)}_{i\to j}}{{N\choose \ell} - E^{(k\to \ell)}_i \choose E^{(\ell)}_j-E^{(k\to \ell)}_{i\to j}}\right\}.   
\end{align}

Putting it all together, we can construct a normalized mutual
information~$\text{NMI}_{\text{cross}}$ that incorporates cross-layer
similarity as follows. For the entropy, we can use the same expression
as in Eq.~(\ref{align-ent}), giving
\begin{align}\label{cross-ent}
    \text{H}_{\text{cross}}(G_i) = \sum_{\ell\in \mathcal{L}}\log {{N\choose \ell} \choose E^{(\ell)}_i}.     
\end{align}
And for the conditional entropy under this encoding, we can modify
Eq.~(\ref{align-CE}) appropriately to account for the best
layer~$k_i^{(\ell)}$ in $G_i$ with which to transmit
layer~$G_j^{(\ell)}$. The resulting expression is
\begin{align}\label{cross-CE}
    \text{H}_{\text{cross}}(G_j\vert G_i)
 = \sum_{\ell \in \mathcal{L}}\log {E^{(k_i^{(\ell)}\to \ell)}_i \choose E^{(k_i^{(\ell)}\to \ell)}_{i\to j}}{{N\choose \ell} - E^{(k_i^{(\ell)}\to \ell)}_i \choose E^{(\ell)}_j-E^{(k_i^{(\ell)}\to \ell)}_{i\to j}}.
\end{align}
As before, subbing the entropy and conditional entropy into
Eq.~(\ref{general-NMI}) gives the NMI under this cross-layer encoding.
The $\text{NMI}_{\text{cross}}$ measure is the most flexible and
nuanced of the measures we present here. As such, it is the primary
NMI measure of focus for the experiments in Sec.~\ref{sec:results}.

It is worth noting that any NMI measure constructed using the
framework in Sec.~\ref{sec:hypergraph-NMI}---and thus the three
measures $\text{NMI}_{\text{bulk}}$, $\text{NMI}_{\text{align}}$,
and $\text{NMI}_{\text{cross}}$ of this section---will give a
maximum score of $1$ for isomorphic hypergraphs $G_1,G_2$ when
their node labels are correctly aligned. This is because, once the
(complete) structural overlap among $G_1,G_2$ is known, we know
everything about $G_2$ after knowing $G_1$, and vice versa.
However, \emph{non-isomorphic} hypergraphs can also obtain the
maximum similarity score of $1$ for $\text{NMI}_{\text{cross}}$
(but not $\text{NMI}_{\text{bulk}}$ or
$\text{NMI}_{\text{align}}$), since similarity is assessed across
layers. Specifically, $\text{NMI}_{\text{cross}}(G_1,G_2)=1$ for
any pair of hypergraphs $G_1,G_2$ for which each of the layers in
one hypergraph are fully nested within at least one layer of the
other hypergraph. This flexibility is crucial for assessing
similarity beyond pure structural isomorphism, allowing for the
nested structures ubiquitous in real hypergraphs to contribute to
their similarity. Moreover, such nestedness is critical for
understanding structural redundancy in higher-order
systems~\cite{ceria2025relevance,landry2024filtering,barrett2025counting}.
In this context, information encodings that account for nested
structures can be used to understand which layers of a hypergraph
are most critical for summarizing its higher-order
structure~\cite{kirkley2025reducibility, lucas2026reducibility}.

In Fig.~\ref{fig:fig1}, we show the results of applying our measures
to three small example hypergraphs on the same set of $N=8$ nodes. We
can see that all three hypergraphs generally share similar structure
across their layers, but that the three measures vary substantially
across all pairs.  Next to each hypergraph pair, for reference, we
plot matrices showing the graph NMI measure
of~\cite{felippe2024network} applied to the projections of each layer
to the lower order of the two.  These order-order similarity matrices
are defined with entries
\begin{align}
    \text{I}_{\ell\ell'} = h_b(p_\ell)+h_b(p_{\ell'})-h_s(\bm{P}_{\ell\ell'}),
\end{align}
where the orders satisfy $\ell'\geq \ell$, $h_b(x)=-p\log p -(1-p)\log
(1-p)$ is the binary entropy, $h_s(\bm{x})=-\sum_{i}x_i\log x_i$ is
the Shannon entropy, and $\bm{P}_{\ell\ell'} =
\{p_{\ell\ell'},p_{\ell}-p_{\ell\ell'},p_{\ell'}-p_{\ell\ell'},1-p_{\ell}-p_{\ell'}+p_{\ell\ell'}\}
$ is a vector totaling the overlaps among the layers (analogous to a
confusion matrix), with
$p_{\ell}=\abs{G^{(\ell)}}/{N\choose \ell}$,
$p_{\ell'}=\abs{G^{(\ell'\to\ell)}}/{N\choose \ell}$, and
\mbox{$p_{\ell\ell'}=\abs{G^{(\ell'\to \ell)}\cap
G^{(\ell)}}/{N\choose \ell}$} the relevant densities of hyperedges
used to compute the entries.

The hypergraphs $G_1$ and $G_2$ are similar in that $G^{(2)}_1$ has a
high structural overlap with $G^{(3)}_2$, and $G^{(3)}_1$ (with its
four 3-hyperedges) has a high structural overlap with the single
4-hyperedge of $G^{(4)}_2$.  However, only $\text{NMI}_{\text{cross}}$
is able to capture this similarity, giving a high value of $0.88$. The
other measures are unable to see any overlap among the hypergraphs
except for the single edge~$(7,8)$, both giving low scores.
Meanwhile, $G_2$ and $G_3$ are similar in that $G^{(3)}_2$ has a high
structural overlap with $G^{(2)}_3$, and $G^{(4)}_2$ has maximal
structural overlap with $G^{(4)}_3$.  Since some of this overlap is
now occurring at the same order~$\ell=4$---i.e., is \emph{intra-order}
similarity---both $\text{NMI}_{\text{bulk}}$ and
$\text{NMI}_{\text{align}}$ are able to detect it, giving moderate
scores.  Looking at Eqs.~(\ref{align-ent}) and~(\ref{align-CE}), we
can see that $\text{NMI}_{\text{align}}$ will scale positively with
both the number of overlapping hyperedges \emph{and} the size of those
hyperedges, contributing to fluctuations in this measure across the
pairs of hypergraphs.  Meanwhile, $\text{NMI}_{\text{cross}}$ still
gives the highest score, detecting both the intra- and cross-order
similarity among the hypergraphs. Finally, $G_1$ and $G_3$ are similar
in that $G^{(2)}_1$ has a high structural overlap with $G^{(2)}_3$,
and $G^{(3)}_1$ has a high structural overlap with $G^{(4)}_3$. In
this case, since most of the hyperedges are of order $\ell=2$, where
$G_1$ and $G_3$ overlap, both $\text{NMI}_{\text{bulk}}$ and
$\text{NMI}_{\text{align}}$ detect relatively high similarity values.
Meanwhile, $\text{NMI}_{\text{cross}}$ returns a perfect similarity
score---$G^{(2)}_1$ and $G^{(2)}_3$ are identical, while $G^{(3)}_1$
is perfectly nested within $G^{(4)}_3$. 

As discussed, $\text{NMI}_{\text{cross}}$ can obtain its maximum
value of $1$ for non-isomorphic hypergraphs, so long as they are
completely nested within each other. However, nestedness must
occur on a layer-layer basis to obtain perfect similarity. Thus,
for $G_2$ and $G_3$---for which $G^{(2)}_3$ is nested within the
union of layers $G^{(2)}_2$ and $G^{(3)}_2$---we have a slight
information penalty since $G^{(2)}_3$ is not perfectly nested
in either $G^{(2)}_2$ or $G^{(3)}_2$ individually (if multiple
layers of $G_2$ could be used to transmit a single layer of
$G_3$, one would incur an additional information cost for
specifying the combination of layers, as well as a potentially
substantial computational cost for checking possible layer
combinations).

Finally, we remark that one can apply the idea of cross-order
similarity to the different orders of the same hypergraph in order
to capture redundancies and reduce the dimensionality of the
system preserving its essential structural
features~\cite{kirkley2025reducibility}.

\begin{figure*}[ht]
\includegraphics[width=1.0\textwidth]{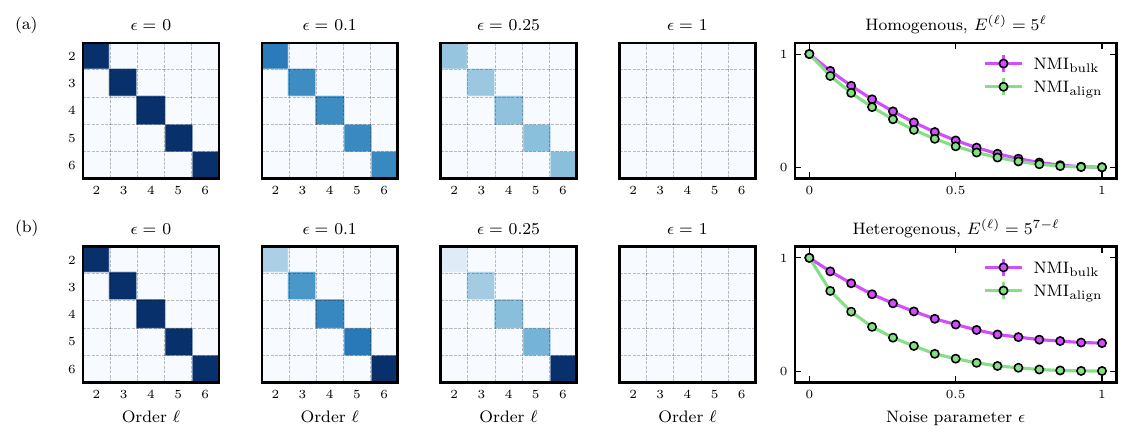}
\caption{
    \textbf{Information theory measures for intra-order hypergraph similarity.}
    (\textbf{A})~Random hypergraphs with homogeneous layer densities.
    Order-order graph NMI values for the layers' pairwise projections
    (left) show maximum shared structure at $\epsilon=0$, which
    decreases uniformly as the layers are randomized. The intra-order
    hypergraph similarity measures $\text{NMI}_{\rm bulk}$ and
    $\text{NMI}_{\rm align}$ smoothly decrease with the noise
    $\epsilon$, reaching zero in the regime of complete noise (right).
    Due to the homogeneous hyperedge densities across layers, both NMI
    measures give similar values.
    (textbf{B})~Random hypergraphs with heterogeneous layer densities.
    Order-order similarities (left) indicate higher intra-order
    similarity for larger orders as noise is applied, due to the
    heterogeneous densities of the layers.
    In this case, \textcolor{black}{on the right-hand plot}, we see
    that $\text{NMI}_{\rm bulk}$ inflates the mutual information
    contributions for high $\epsilon$, resulting in a non-negligible
    NMI value at $\epsilon=1$. The $\text{NMI}_{\rm align}$ measure
    does not have this issue, vanishing in the high noise regime.  }
\label{fig:bulkalign}
\end{figure*}

Numerically, $\text{NMI}_{\text{bulk}}$ and
$\text{NMI}_{\text{align}}$ can be computed quickly with a runtime
linear in the number of hyperedges in $G_1$ and $G_2$ by using
set~overlaps. However, $\text{NMI}_{\text{cross}}$ is more challenging
to compute due to the computation of the overlap $E^{(k\to
\ell)}_{i\to j}$ in Eq.~(\ref{overlap-cross}) and the projected layer
size $E^{(k\to \ell)}_{i}$ in Eq.~(\ref{cross-CE}). This is because
direct projection of the layer~$G^{(k)}_i$ to obtain $G^{(k\to
\ell)}_i$ quickly becomes computationally intractable as $k,\ell$
become large. For example, when $k=30$ and $\ell=15$, each hyperedge
in layer~$k$ has over $100$-million sub-tuples of size~$\ell$ which we
must project onto to obtain the unique hyperedges contributing to
$G^{(k\to \ell)}_i$. In the SM~\ref{app:algorithm}, we describe a
recursive algorithm to implement $\text{NMI}_{\text{cross}}$
efficiently, allowing for the fast comparison of hypergraphs with
millions of nodes and large hyperedge orders using our NMI measures.

In the SM~\ref{app:multiscale}, we describe how to extend our measures
to quantify the shared information among arbitrary coarse-grainings of
nodes between a pair of hypergraphs. These \emph{multiscale}
hypergraph NMI measures allow for capturing hypergraph similarity at
the scale of interest, as well as adapting the measures to
multigraphs. 

\section{Results}
\label{sec:results}

To illustrate the hypergraph similarity measures introduced above, we
first examine systems with variable intra-order similarity using the
${\rm NMI}_{\rm bulk}$ and ${\rm NMI}_{\rm align}$ measures. We then
move through the hierarchy of measures and study hypergraphs with
variable cross-order similarity, showing that the ${\rm NMI}_{\rm
cross}$ measure---the most expressive and flexible measure developed
in our hierarchy of measures---more adequately captures such
similarity than ${\rm NMI}_{\rm align}$.  Finally, we apply ${\rm
NMI}_{\rm cross}$ to three empirical hypergraphs representing
collaboration patterns in physics, the film industry, and software
development, to analyze the patterns that are revealed in these
systems using our framework.

\subsection{Intra-order similarity}

To control the level of intra-order similarity among pairs of
hypergraphs, we generate an initial hypergraph~$G_1$ as a random
hypergraph over $N=100$ nodes in which each layer $G_1^{(\ell)}$ for
$\ell \in \{2,3,4,5,6\}$ is generated with a fixed number of
hyperedges $E_1^{(\ell)}$ chosen uniformly at random from all
${N\choose \ell}$ possibilities. We then generate a second hypergraph
$G_2$ by starting with a copy of $G_1$ and perturbing the hyperedges
in $G_2$ according to a noise parameter $\epsilon\in [0,1]$. For each
value $\epsilon$, we choose a fraction $\epsilon$ of $G_2$'s
hyperedges uniformly at random and replace each with a randomly chosen
hyperedge of the same size. In this way, for $\epsilon=0$ we have that
$G_1$ and $G_2$ are identical, while at $\epsilon=1$ they are both
equivalent to independently generated random hypergraphs.  We then
compute both $\text{NMI}_{\rm bulk}$ and $\text{NMI}_{\rm align}$ as
we continue to inject structural noise by increasing $\epsilon$. As
discussed in Sec.~\ref{sec:hierarchy}, the $\text{NMI}_{\rm align}$
measure is able to correct for heterogeneous densities across layers,
a feature observed in many real-world
hypergraphs~\cite{cencetti2021temporal}, and which is not accounted
for in $\text{NMI}_{\rm bulk}$. Therefore, we vary the relative
densities $\rho^{(\ell)}=E^{(\ell)}/{N\choose \ell}$ of the layers
in the initial random hypergraph $G_1$ to examine the resulting
discrepancy in the two measures.  

Figure~\ref{fig:bulkalign} shows the results of these experiments.
Each simulation was repeated ten times and the results were averaged,
with error bars (vanishingly small for these experiments) indicating
three standard errors in the mean. In row~(a) we plot the results for
hypergraphs generated with \mbox{$E^{(\ell)}=5^{\ell}$}, to capture
the exponential increase in the number of edges required to maintain a
constant density $\rho^{(\ell)}$ as we increase $\ell$. (Maintaining
$\rho^{(\ell)}$ exactly while keeping a reasonable overall edge count
results in too extreme a level of heterogeneity in the distribution of
edge counts across layers, with \mbox{$E^{(\ell)}\approx 0$} for all
$\ell$ lower than the highest order.) The left four columns show the
order-order similarity matrices computed using the graph NMI measure
of~\cite{felippe2024network} applied to the projections of each layer
to the lower order of the two. We can see that the density of edges
within each layer is unchanged by the noise, and that the overlaps are
quite homogeneous across the diagonal of the matrices due to the
homogeneous layer densities. The off-diagonal entries nearly vanish in
all cases, due to the overall sparsity of the hypergraphs and lack of
nestedness among the layers. In the rightmost column we show the
results of applying our NMI measures. We see that both have a smooth
decrease with the injected noise, as expected, reaching zero for
$\epsilon=1$. This illustrates that, for homogeneous edge densities
across layers, both NMI formulations are capable of distinguishing
meaningful hypergraph overlap from the spurious overlap expected due
to the edge density. 

However, this scenario---higher hyperedge counts for higher
orders~$\ell$---is unlikely to be observed in practice. It is instead
more realistic in practice to observe \emph{lower} edge counts as we
increase the
order~$\ell$~\cite{battiston2020networks,cencetti2021temporal}. In
row~(b) we show the same experiments for decreasing layer sizes
\mbox{$E^{(\ell)}=5^{7-\ell}$}---this form ensures that the total
number of edges, hence overall edge density~\mbox{$E/(2^N-N-1)$}, is
the same as in the previous experiments---which tell a different
story. We can see that in this case, the similarity matrices to the
left indicate a high level of heterogeneity in the similarities
between layers. This results in very little change to
$\text{NMI}_{\text{align}}$ in the rightmost panel as we add noise to
the system, with the curve approaching zero as before. However, the
heterogeneous layer densities result in a severely inflated value of
$\text{NMI}_{\text{bulk}}$ in the high noise regime, meaning it is no
longer able to correct for the spurious overlap we see based on the
densities of the layers. This suggests, as described in
Sec.~\ref{sec:hierarchy}, that the $\text{NMI}_{\text{align}}$ measure
is more appropriate for capturing intra-order similarity among
hypergraphs with heterogeneous edge densities across layers.  

\begin{figure}[t]
\includegraphics[width=0.50\textwidth]{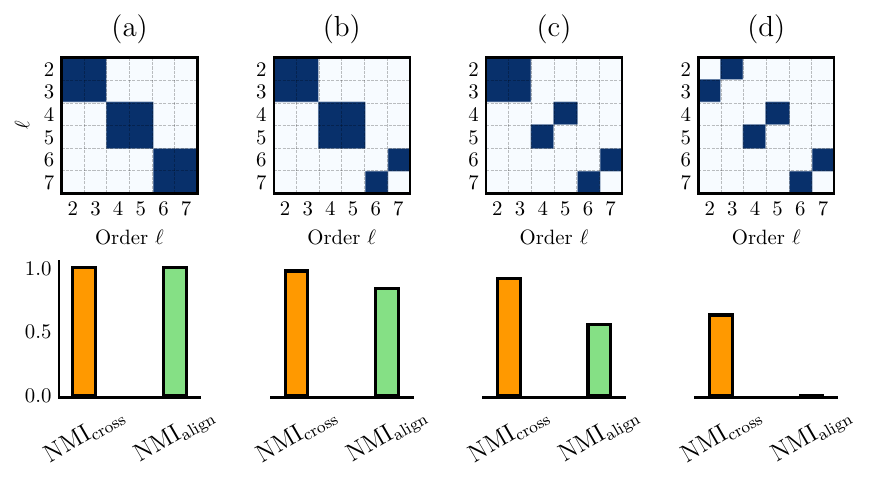}
\caption{
    \textbf{Information theory measures for cross-order hypergraph similarity.}
    (\textbf{A}) Two initial random hypergraphs share the same layers
    $\ell=2$, 4, and 6, which are nested inside of $\ell=3$, 5, and 7,
    respectively.
    (\textbf{B}) Layers 6 and 7 are perturbed, causing their
    respective blocks to lose intra-order similarity. The intra-order
    measure $\text{NMI}_{\text{align}}$ is thus reduced, while the
    cross-order measure $\text{NMI}_{\text{cross}}$ changes
    negligibly.
    (\textbf{C}) Layers 4 and 5 are further perturbed, removing
    another nested block.  Intra-order similarity is significantly
    reduced yet again.
    (\textbf{D}) Finally, layers 2 and 3 are perturbed, dismantling
    all blocks and eliminating any similarity between layers of equal
    size. The intra-order score $\text{NMI}_{\rm align}$ approaches
    the minimum value of zero, while $\text{NMI}_{\rm cross}$ is still
    able to capture the structural similarity across different orders
    of interaction.
}
\label{fig:cross}
\end{figure}

\subsection{Cross-order similarity}

While the intra-order comparisons made by $\text{NMI}_{\text{bulk}}$
and $\text{NMI}_{\text{align}}$ are relevant for cases where different
orders of interaction are considered independent from one another, in
many real-world applications it is important to understand the
structural similarity among hypergraphs while accounting for the
nestedness of these interactions, as interactions that are nested may
influence each other~\cite{battiston2021physics}. It is therefore
important to understand cross-order similarity among hypergraphs,
which is the strength of the proposed $\text{NMI}_{\text{cross}}$
measure.   

To understand how $\text{NMI}_{\text{cross}}$ performs compared to the
intra-order similarity measure $\text{NMI}_{\text{align}}$ when
nestedness is perturbed, we design a third generative model, the
block-nested hypergraph, which explicitly encodes dependencies between
layers of different sizes across hypergraphs. In this model, we first
generate ``parent'' layers $\ell\in\{3,5,7\}$ in $G_1$ and $G_2$ as
random hypergraphs on $N=100$ nodes with $E^{(\ell)}=N\binom{7}{\ell}$
hyperedges for each~$\ell$.  We then pick the layers  $\ell=2$, 4, and
6 to be ``child'' layers, corresponding to the parent layers $\ell=3$,
5, and 7, respectively. We leave the parent layers unchanged and set
each child layer $G^{(\ell)}_j$ in hypergraph $G_j$ to be the
projection $G^{(k)}_i$ of its parent layer in the other hypergraph
$G_i$. This creates hypergraphs with perfect cross-order overlap among
parent-child layer pairs across the hypergraphs. To vary the level of
intra-order similarity, we keep some parent layers $\ell\in \{3,5,7\}$
identical across the two hypergraphs and allow others to be generated
independently at random.

We show the results of applying $\text{NMI}_{\text{align}}$ and
$\text{NMI}_{\text{cross}}$ to these synthetic hypergraph pairs in
Fig.~\ref{fig:cross}. In the top row we plot the order-order
similarity (as in Fig.~\ref{fig:bulkalign}) across the two hypergraphs
for each experimental setting, and in the bottom row we plot the two
NMI measures as a bar chart. In column~(a) we keep all parent
layers $\ell\in\{3,5,7\}$ identical across the hypergraphs. In
this case, both measures return a value of $1$ as expected.
\textcolor{black}{In the system shown in column (b), the parent
layer $\ell=7$ is generated independently at random across the two
hypergraphs.  However, layer $\ell=6$ of $G_2$ is still generated
as a ``child'' nested within layer $\ell=7$ of $G_1$, while layer
$\ell=6$ of $G_1$ is the child of layer $\ell=7$ of $G_2$.  Thus,
generating the layers $\ell=7$ independently destroys the
\emph{intra-order} similarity at $\ell=7$, but the parent-child
relationship still enforces overlap between layers $\ell=7$ and
$\ell=6$ \emph{across} the two hypergraphs.} We can see that this
perturbation results in nearly no detectable change in
$\text{NMI}_{\text{cross}}$ and a moderate decrease in
$\text{NMI}_{\text{align}}$. 

In the system of column (c), we proceed from the configuration in
column (b) and generate layer $\ell=5$ independently at random
across hypergraphs. Again, by construction, layer $\ell=4$ of
$G_2$ is generated to be nested within layer $\ell=5$ of $G_1$,
and similarly $\ell=4$ of $G_1$ is nested within layer $\ell=5$ of
$G_2$.  In this case, we can see again that the cross-order
measure $\text{NMI}_{\text{cross}}$ is nearly unchanged, while
$\text{NMI}_{\text{align}}$ exhibits a notable decline to around
$0.5$. Finally, in column~(d) we fully destroy the intra-order
similarity by allowing all parent layers to be generated
independently while preserving the parent-child relationships
across layers. Here we can see that $\text{NMI}_{\text{cross}}$
shows a modest drop, while $\text{NMI}_{\text{align}}$ almost
completely vanishes. Drops in $\text{NMI}_{\text{cross}}$ are due
to the transmission cost of the parent layers~$\ell=3,5,7$: since
these do not have any parent from which they can be transmitted
cheaply, they must be transmitted from layers in the opposite
hypergraph that are potentially uncorrelated, decreasing the NMI.

In the SM~\ref{section:similaritynestedness}, we further examine the
three proposed NMI measures in various other models of hypergraphs
with tunable nestedness, finding intuitive results that show little
discrepancy for non-nested systems and support the usage of
$\text{NMI}_{\text{cross}}$ for nested systems. 

\begin{figure*}[t]
\includegraphics[width=1.0\textwidth]{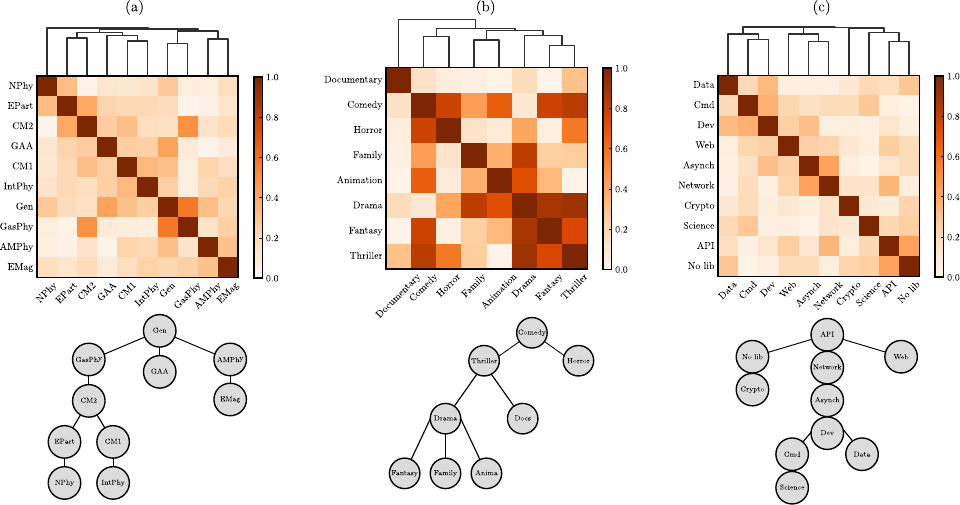}
\caption{
    \textbf{Hypergraph similarity for real-world systems.}  
    NMI matrices among all pairs of hypergraphs within real-world
    systems arising across various disciplines (top row), each
    accompanied by its corresponding minimum spanning tree using
    $1-\text{NMI}$ as an edge weight.
    (\textbf{A})~Multiplex hypergraph of co-authorship among physics
    authors in different physics fields.    
    (\textbf{B})~Multiplex hypergraph of co-appearances among actors
    in different film genres. 
    (\textbf{C})~Multiplex hypergraph of repository co-editing among
    software development teams.  For each system, the similarity among
    hypergraphs corresponding to qualitatively similar subjects
    (e.g.~nuclear and elementary particle physics) tend to be higher.
}
\label{fig:empirical}
\end{figure*}

\subsection{Applications to real-world systems}

To illustrate the applicability of our information theoretic framework
for hypergraph similarity, we study three empirical systems that are
naturally represented as \textcolor{black}{\emph{multiplex}
hypergraphs---systems consisting of multiple independent
hypergraphs on the same set of nodes.} We study collaboration
networks from three different disciplines:
physics~\cite{pacs_data}, film~\cite{lotito2024multiplex}, and
software
development~\cite{schueller2022evolving,schueller2024modeling,betti2025dynamics}.
In each dataset, a hyperedge is formed among $\ell$ nodes
(individuals) if these $\ell$ individuals contributed to the same
paper/movie/repository~\cite{lotito2023hypergraphx,lotito2026hypergraphxdata}.
The hypergraphs in each multiplex system are organized by
categorical metadata: subfields of physics in the American Physical Society (APS) dataset,
movie genres in the Internet Movie Database (IMDb) corpus, and repository tags in the Rust
open-source ecosystem. Further descriptions and summary statistics
for these datasets can be found in the SM~\ref{section:multiplex}.

Figure~\ref{fig:empirical} shows the results of applying the
$\text{NMI}_{\text{cross}}$ measure to all pairs of
\textcolor{black}{hypergraphs in each multiplex}. In the top row, we
show matrices containing these hypergraph NMI values, along with
corresponding dendrograms constructed from hierarchical clustering
using the Ward criterion. Below each similarity matrix, we plot a
minimum spanning tree (MST) constructed using $1-{\rm NMI}_{\rm
cross}$ as edge weights. For each dataset, the NMI values between
hypergraphs from qualitatively similar categories of interactions tend
to be high, with less similarity assigned to disparate categories.
For example, in the APS dataset, we see a high value of similarity
between Nuclear (NPhy) and Elementary Particle physics (EPart), which
in turn are dissimilar to Gas physics (GasPhy). Meanwhile, for the
IMDb dataset we observe high similarity among the Thriller and Drama
genres, for which the genre boundary is often unclear. Meanwhile, we
see very little similarity among documentaries and other genres due to
the fundamentally different nature of acting in documentaries.
Finally, for open-source software collaborations we see a high NMI
between the hypergraphs corresponding to command line utilities (Cmd),
development tools (Dev), and data structures (Data), which are
functionally related via shared code and common use of custom data
types.  Section~\ref{section:multiplex} of the SM shows the
order-order similarity matrices among pairs of
\textcolor{black}{hypergraphs in} each dataset, giving support to the
findings discussed here. And in Section~\ref{section:runtime} we show
the runtime scaling of our measures on these empirical datasets.

In Section~\ref{section:temporal} of the SM we explore a further
application of our method to real hypergraph data, this time to detect
anomalies in temporal higher-order interactions.  \textcolor{black}{In
Section~\ref{section:dynamcis} we then study the sensitivity of
$\text{NMI}{\rm cross}$ for detecting structural overlaps that impact
dynamical contagion processes on hypergraphs, connecting this measure
of structural similarity to similarity in dynamical behaviors.
Finally, in Section~\ref{section:comparison} we compare our measures
with a simple Jaccard similarity baseline and alternative measures of
structural similarity~\cite{agostinelli2025higher}.} These results,
together with those in our synthetic tests, show the effectiveness of
our hypergraph similarity measures for capturing meaningful structural
overlap among hypergraphs in both real and controlled settings.

\section{Discussion}

Given the growth in complexity and dimensionality of relational
datasets, quantifying the similarity between hypergraphs is an
increasingly important challenge in network science. Existing
approaches to this task tend to rely on ad hoc heuristics and/or
tunable parameters, to which results are highly sensitive. Moreover,
many of these methods fail to scale to large real-world hypergraphs
with varying layers of higher-order interactions, with no prescription
for correcting for spurious overlaps due to edge density. In this
work, we introduce a principled, flexible framework for
constructing mutual information measures between hypergraphs, and
construct a hierarchy of hypergraph similarity measures using this
framework to highlight structural overlaps among hypergraphs at
different scales and orders of interaction. Across a series of
synthetic experiments, we showed that each measure behaves in an
intuitive manner, highlighting structural similarity precisely in
the way prescribed by the measure's encoding scheme. In
particular, our measure $\text{NMI}_{\text{cross}}$ proved
essential to capture the most general notion of hypergraph
similarity, detecting structural correspondence both within and
across layers of different orders. Meanwhile, extension to a
multiscale measure was required for detecting similarity beyond
the node-level. We further demonstrated the practical value of our
methods through applications to empirical multiplex hypergraphs
from three distinct collaboration domains.

The proposed framework opens up several avenues for future
exploration. We only explored a few different encodings in this
paper---the bulk, align, and cross encodings, along with their
corresponding multiscale variants. However, the  NMI measure in
Eq.~(\ref{general-NMI}) applies to any desired encoding, leaving open
the possibility for more sophisticated models capturing more nuanced
aspects of similarity among hypergraphs. For example,
in~\cite{felippe2024network} we explore a degree-corrected variant of
the graph NMI to capture ego network-level similarities, a variant
which would be  possible to explore also for higher-order data.
Moreover, one could explore compression via community
structure~\cite{peixoto2023implicit} to more efficiently encode the
hypergraphs being compared. Our method might also be extended to
account for node or edge metadata, as well as temporal or multiplex
hypergraph structure, to accommodate a wider variety of real-world
data. Finally, it may be possible to modify these measures to encode
and cluster entire populations of more than two hypergraphs arising
from longitudinal or cross-sectional studies, extending existing work
for pairwise
graphs~\cite{kirkley2024inference,kirkley2023compressing}.

There are also numerous important applications in which our framework
might prove useful. In neuroscience, for example, higher-order network
representations of neural activity also provide distinct structural
signatures undetectable with standard pairwise
methods~\cite{santoro2024higher,neri2025taxonomy}. Our framework would
allow for the comparison and clustering of higher-order brain networks
across subjects and experimental conditions in such studies to reveal
underlying regularities in functional connectivity.

Finally, there are a few key limitations of the hierarchy of NMI
measures we present here which, if addressed, can allow for the
proposed NMI framework to be applied to a wider variety of
systems. Firstly, although the NMI framework in
Sec.~\ref{sec:hypergraph-NMI} applies to any pair of hypergraphs,
the proposed hierarchy of NMI measures in Sec.~\ref{sec:hierarchy}
apply only to node-aligned hypergraphs. This covers a wide range
of cross-sectional and longitudinal network measurements but is
not inclusive of all possible hypergraphs one may wish to compare.
For hypergraphs with the same number of nodes but without
alignment in their labels, one can in principle fix the node
labels of $G_1$ and sample the space of node labelings for $G_2$
with simulated annealing or another MCMC technique to find the
configuration that maximizes the NMI between the two hypergraphs.
Meanwhile, for unaligned networks with different numbers of nodes,
one could expand this scheme for node alignment by restricting the
graph with more nodes to a subset of nodes equal in size to other
hypergraph; finding the best alignment with simulated annealing as
described above; and repeating these two steps until convergence
of the node labels and NMI (alternatively, one could add nodes to
the smaller of the two hypergraphs until the number of nodes is
equal, then apply the stochastic matching without the need for
selecting a subset of nodes). Both of these modifications would,
however, incur a substantial additional computational expense to
the otherwise highly efficient NMI measures presented here.
Additionally, while the mesoscale NMI we propose can be directly
applied to compare hypergraphs with edge weights that are positive
integers (see SM~\ref{app:multiscale}), our measures do not yet
account for hypergraphs with non-integer or negative edge weights.
Such an extension could increase the compatibility of our
framework with hypergraphs constructed from time series data, in
which correlations may be both signed and continuous.

Our work provides foundational tools for the principled comparison of
higher-order network datasets, shedding light on the structural
organization of empirical systems with non-dyadic interactions.

\section*{Acknowledgments}
\vspace{-\baselineskip}
The authors acknowledge F.~Malizia for insightful discussions on
spreading dynamics.

\subsection*{Funding}
F.B.~acknowledges support from the Austrian Science Fund~(FWF) through
projects~10.55776/PAT1052824 and 10.55776/PAT1652425.
A.K.~acknowledges support from the HKU-100 Start Up Fund and the
National Science Foundation of China through the Young Scientist Fund
Project No.~12405044.

\subsection*{Author contributions} 
A.K. developed the methodology. H.F. and A.K. wrote the original
draft. H.F. and A.K. developed the software, curated the data, and
created the visualizations. F.B. and A.K. supervised the project. 
H.F., A.K., and F.B. contributed to the investigation, formal analysis, validation, and project administration. 
H.F., A.K., and F.B. reviewed and edited the
manuscript.

\subsection*{Competing interests} 
The authors declare no competing interests.

\subsection*{Data, Code, and materials availability}
Code and data are available as part of the library Hypergraphx
at~\url{https://doi.org/10.1093/comnet/cnad019}.
All data and code needed to evaluate and reproduce the results in the
paper are present in the paper and/or the Supplemental Materials. 
sciadvs required This study did not generate new materials. 
required

    %
    %

    \clearpage

    \onecolumngrid

    \begin{center}
      \textbf{\large Supplemental Material for: \\ \vspace{0.25cm}
    Information theory for hypergraph similarity \vspace{0.25cm}} \\[.2cm]

      Helcio Felippe,$^1$ \, Alec Kirkley,$^{1,2,3}$ and Federico Battiston$^{4,5}$, \\ [.1cm]
      {\itshape ${}^1$Department of Network and Data Science, Central University, 1100 Vienna, Austria \\
                ${}^2$Institute of Data Science, University of Hong Kong, Hong Kong SAR, China \\
                ${}^3$Department of Urban Planning and Design, University of Hong Kong, Hong Kong SAR, China \\
                ${}^4$Urban Systems Institute, University of Hong Kong, Hong Kong SAR, China \\
                ${}^5$Department of AI, Data and Decision Sciences, Luiss University of Rome, Rome, Italy \\
                }
    \end{center}

    \setcounter{equation}{0}
    \setcounter{figure}{0}
    \setcounter{table}{0}
    \setcounter{page}{1}
    \setcounter{section}{0}
    \renewcommand{\theequation}{S\arabic{equation}}
    \renewcommand{\thefigure}{S\arabic{figure}}
    \renewcommand{\thetable}{S\arabic{table}}
    \renewcommand{\thepage}{S\arabic{page}}
    \renewcommand{\thesection}{S\arabic{section}}
    \renewcommand{\bibnumfmt}[1]{[#1]}
    \renewcommand{\citenumfont}[1]{#1}

%
%
%

    \section{Efficient implementation of $\text{NMI}_{\text{cross}}$}\label{app:algorithm}

    To calculate $\text{NMI}_{\text{cross}}(G_1,G_2)$ numerically for
    hypergraphs $G_1,G_2$ with large hyperedge orders $\mathcal{L}$, the
    computational bottleneck lies in computing $E_{i\to j}^{(k\to \ell)}$
    in Eq.~(\ref{overlap-cross}) and $E_{i}^{(k\to \ell)}$ in
    Eq.~(\ref{cross-CE}). For $k,l\lesssim 10$, these two quantities can
    be computed by projecting $G^{(k)}_i$ to obtain $G^{(k\to \ell)}_i$,
    then computing the set intersection of $G^{(k\to \ell)}_i$ and
    $G^{(\ell)}_j$ to calculate $E_{i\to j}^{(k\to \ell)}$ and the size of
    $G^{(k\to \ell)}_i$ to calculate $E^{(k\to \ell)}_i$. However, for
    $k,l\gtrsim 10$, the direct projection of a $k$-tuple onto its
    ${k\choose \ell}$ subsets of size~$\ell$ becomes computationally
    costly, effectively becoming intractable for $k,l\gtrsim 30$. Here we
    design a recursive counting scheme to determine $E_{i\to j}^{(k\to
    \ell)}$ and $E^{(k\to \ell)}_i$ directly without projection, allowing
    us to compute $\text{NMI}_{\text{cross}}(G_1,G_2)$ efficiently for
    hypergraphs with large hyperedges.

    \medskip

    For large $k,\ell$, we can compute $E_{i}^{(k\to \ell)}$ by iterating
    through the edges~$G_i^{(k)}=\{e_1,\dots,e_{E_i^{(k)}}\}$ in a fixed
    order, for each edge~$e_t$ checking its overlaps
    \mbox{$o(e_t)=\{e_t\cap e_\tau:\tau < t\}$} with all previously
    checked edges. Then, we can compute the number of new projected tuples
    that $e_t$ contributes to $E_i^{(k\to \ell)}$ as ${k\choose
    \ell}-E^{(o(e_t)\to \ell)}$, where $E^{(o(e_t)\to \ell)}$ is the
    number of unique subtuples of size~$\ell$ within the set of
    overlapping tuples~$o(e_t)$, which can be computed recursively using
    the same approach. Meanwhile, the overlap $E_{i\to j}^{(k\cap\ell)}$
    can be efficiently computed by iterating over the hyperedges~$e_t\in
    G_i^{(k)}$ and incrementing $E_{i\to j}^{(k\cap \ell)}$ for each
    edge~$e_s\in G_j^{(\ell)}$ that fully overlaps with the larger
    tuple~$e_t$, removing $e_s$ from $G^{(\ell)}$ after the comparison if
    it overlapped with $e_t$. 

    \medskip

    The computations of $E_{i}^{(k\to \ell)}$ and $E_{i\to j}^{(k\to
    \ell)}$ using the above counting methods incur a total computational
    complexity of roughly
    $O\left[(E_i^{(k)})^2+E_i^{(k)}E_j^{(\ell)}\right]$ rather than the
    $O\left({k\choose \ell}E_i^{(k)}\right)$ complexity using the
    projection~$k\to \ell$. Thus, it becomes more efficient to use these
    algorithms for ${k\choose \ell}\gtrsim E_i^{(k)},E_j^{(\ell)}$. Since
    the conditional entropy is computed for all layer pairs~$k\geq \ell$
    in hypergraphs $i,j$ respectively to determine $k_i^{(\ell)}$ in
    Eq.~(\ref{best-layer}), the overall runtime complexity for computing
    the $\text{NMI}_{\text{cross}}(G_1,G_2)$ is at most roughly
    $O(E^2L^2)$, with $E$ the typical number of hyperedges in any given
    layer~$\ell$. In practice, we find that this measure easily scales to
    hypergraphs with millions of nodes and hundreds of layers.

    \section{Multiscale hypergraph similarity}\label{app:multiscale}

    The encodings of Sec.~\ref{sec:hierarchy} explore various ways to
    compute similarity among hypergraphs at the node-level, meaning that
    two hyperedges in different hypergraphs only contribute shared
    information to the NMI if they have exactly the same node set or one
    is a subset of the other. But in some applications it is more relevant
    to assess the similarity among two systems at a coarser scale, beyond
    the node-level. For example, when examining whether two hypergraphs
    have statistically similar modular structure---which, crucially, does
    not necessarily mean overlap among their individual hyperedges---the
    measures of Sec.~\ref{sec:hierarchy} fail to capture the desired
    aspects of similarity. To consider an extreme example, a pair of
    hypergraphs generated from the exact same ensemble of sparse
    hypergraphs with identical node community
    partitions~\cite{ruggeri2024framework} will have almost no overlap
    according to the measures of Sec.~\ref{sec:hierarchy}, and thus will
    have NMI scores close to zero. These networks are statistically
    identical at the level of their modular structure, by construction,
    but one must ``zoom out'' beyond the node-scale to the community-scale
    to capture it.

    \medskip

    Following a similar line of reasoning as in~\cite{felippe2024network},
    we can generalize the NMI measures of Sec.~\ref{sec:hierarchy} to a
    family of \emph{multiscale} NMI measures that assess similarity among
    the pair of hypergraphs~$G_1,G_2$ with respect to a shared
    partition~$\bm{b}$ of their nodes, where $b_n$ is the label of
    node~$n\in \{1,\dots,N\}$. The partition~$\bm{b}$ of the nodes can be
    obtained either exogenously from node metadata or endogenously based
    on network structure, for example a community detection algorithm. In
    this case, we do not want to compare the similarity among $G_1$ and
    $G_2$ directly, but rather coarse-grained versions~$\tilde
    G^{(\bm{b})}_1$ and $\tilde G^{(\bm{b})}_2$ of these hypergraphs in
    which all nodes of the same group label in $\bm{b}$ are treated as
    identical. The object $\tilde G^{(\bm{b})}_i$ can be mathematically
    treated as a \emph{multiset} in which each $\ell$-tuple
    (edge)~$(n_1,\dots,n_\ell)$ in layer~$\ell$ is converted to an
    $\ell$-tuple~\mbox{$(b_{n_1},\dots,b_{n_\ell})$} of partition
    labels---sorted to correctly account for duplicates---and identical
    tuples may be repeated. Letting $B$ be the number of unique node
    labels (e.g. groups) in $\bm{b}$, and the scale of individual nodes to
    be $O(N^{-1})$, the multiscale similarity measures we propose assess
    similarity between $G_1$ and $G_2$ at the scale $O(B^{-1})$. Thus,
    when we have few groups, $B\sim O(1)$, our multiscale NMI measures
    assesses hypergraph similarity at the macro-scale~$O(1)$. On the other
    hand, when $B\sim O(N)$ and we have an extensive number of small
    groups of nodes, our multiscale NMI measures assesses hypergraph
    similarity at the node-scale~$O(N^{-1})$ just as with the measures in
    Sec.~\ref{sec:hierarchy}. In the extreme case~$B=N$, our multiscale
    measures can be used to extend the measures of
    Sec.~\ref{sec:hierarchy} to multigraphs or integer-weighted graphs, as
    these can be represented as multisets on $N$ nodes.

    \medskip

    The multiscale NMI measures are largely the same structurally as the
    standard hypergraph NMI measures we present. However, in the
    multiscale case there are a different number of unique (sorted)
    hyperedges of size~$\ell$ that can be constructed from the $B$ unique
    node labels in $\bm{b}$, which will impact the entropy and conditional
    entropy measures' configuration spaces. In order to adapt our NMI
    measures to compare the multisets~$\tilde G^{(\bm{b})}_1$ and $\tilde
    G^{(\bm{b})}_2$, we need to utilize the \emph{multiset coefficient}
    \begin{align}
        \multiset{n}{k} = {n+k-1\choose k},    
    \end{align}
    which is the number of unique multisets of size $k$ that can be
    constructed from a set of $n$ unique items~\cite{roberts2024applied}.
    Additionally, it will be important to extend the concept of
    intersection to multisets, which can be done by defining the
    intersection~$\cap_m$ of the multisets~$M_1$ and $M_2$ as
    \begin{align}
        M_1 \cap_m M_2 = \sum_{x\in M_1,M_2} \text{min}(M_1(x),M_2(x)),
    \end{align}
    where $M_i(x)$ is the number of times element~$x$ occurs in
    multiset~$M_i$. This reduces to the standard set intersection when
    $M_i(x)\in \{0,1\}$.

    \medskip

    For the multiscale bulk NMI measure, the entropy can be modified as
    follows. There are $\multiset{B}{\ell}={B+\ell-1\choose \ell}$ unique
    undirected hyperedges of size~$\ell$ that can be constructed in
    layer~$\ell$. Therefore, there are
    \begin{align}
        \sum_{\ell=2}^{N}{B+\ell-1\choose \ell} =
        \sum_{\ell=0}^{N}{(B-1)+\ell\choose (B-1)} - B - 1
        = {B+N\choose B} - B - 1
    \end{align}
    ways to construct hyperedges of size up to $N$ using the $B$ unique
    node labels, from which we must choose a multiset of size~$E_i$ to
    specify $\tilde G^{(\bm{b})}_i$. The appropriate modification of
    Eq.~(\ref{bulk-ent}) is then
    \begin{align}\label{multiscale-bulk-ent}
        \text{H}^{(\bm{b})}_{\text{bulk}}(G_i) = \log \multiset{{B+N\choose B} - B - 1}{E_i}.
    \end{align}
    \medskip
    The multiscale bulk conditional entropy measure can then be adapted as
    follows. There are $E^{(\ell)}_i$ hyperedges in layer~$\ell$ of
    $\tilde G^{(\bm{b})}_i$, of which we must choose 
    \begin{align}
        E^{(\bm{b})}_{12} = \abs{\tilde G^{(\bm{b})}_1 \cap_m \tilde G^{(\bm{b})}_2}    
    \end{align}
    hyperedges to specify the hyperedges that overlap with $\tilde
    G^{(\bm{b})}_j$. We then must specify a multiset of
    size~$E_j-E^{(\bm{b})}_{12}$ from the ${B+N\choose B} - B - 1$
    possible hyperedges to specify the remaining hyperedges of $\tilde
    G^{(\bm{b})}_j$. The appropriate modification of the conditional
    entropy is thus   
    \begin{align}\label{multiscale-bulk-CE}
        \text{H}^{(\bm{b})}_{\text{bulk}}(G_j\vert G_i)
        = \log {E_i \choose E^{(\bm{b})}_{12}} \multiset{{B+N\choose B} - B - 1}{E_j-E^{(\bm{b})}_{12}}.
    \end{align}
    As we cannot in general say that
    \mbox{$\text{H}^{(\bm{b})}_{\text{bulk}}(G_j\vert G_i)\leq
    \text{H}^{(\bm{b})}_{\text{bulk}}(G_j)$}, to ensure non-negativity of
    the NMI we enforce the entropy as an upper cutoff to the conditional
    entropy so that \mbox{$\text{H}^{(\bm{b})}_{\text{bulk}}(G_j\vert
    G_i)\to \text{min}\left[\text{H}^{(\bm{b})}_{\text{bulk}}(G_j\vert
    G_i),\,\text{H}^{(\bm{b})}_{\text{bulk}}(G_j)\right]$}. This is
    equivalent to saying that $G_j$ will be transmitted by itself if $G_i$
    does not aid in its transmission, and is a result of the expression
    for the conditional entropy being only an upper bound for this
    multiscale case.
    Equations~(\ref{multiscale-bulk-ent})~and~(\ref{multiscale-bulk-CE})
    can then be plugged into Eq.~(\ref{general-NMI}) to find the
    multiscale variant~$\text{NMI}^{(\bm{b})}_{\text{bulk}}$ of
    $\text{NMI}_{\text{bulk}}$, which is bounded in $[0,1]$.

    \medskip

    Using the same line of logic, we can compute
    $\text{NMI}^{(\bm{b})}_{\text{align}}$ and
    $\text{NMI}^{(\bm{b})}_{\text{cross}}$ using the following adaptations
    of the entropy and conditional entropy measures of
    Sec.~\ref{sec:hierarchy}:
    \begin{align}
        \text{H}^{(\bm{b})}_{\text{align}}(G_i) &= \sum_{\ell\in \mathcal{L}}\log \multiset{\multiset{B}{\ell}}{E^{(\ell)}_i} ,\\
        \text{H}^{(\bm{b})}_{\text{align}}(G_j\vert G_i) &= \sum_{\ell\in \mathcal{L}}\log {E^{(\ell)}_i \choose E^{(\ell,\bm{b})}_{12}}\multiset{\multiset{B}{\ell}}{E^{(\ell)}_j-E^{(\ell,\bm{b})}_{12}},\\
        \text{H}^{(\bm{b})}_{\text{cross}}(G_i) &= \sum_{\ell\in \mathcal{L}}\log \multiset{\multiset{B}{\ell}}{E^{(\ell)}_i},\\
        \text{H}^{(\bm{b})}_{\text{cross}}(G_j\vert G_i) &= \sum_{\ell \in \mathcal{L}}\log {E^{(k_i^{(\ell,\bm{b})}\to \ell)}_i \choose E^{(k_i^{(\ell,\bm{b})}\to \ell,\bm{b})}_{i\to j}}\multiset{\multiset{B}{\ell} }{ E^{(\ell)}_j-E^{(k_i^{(\ell,\bm{b})}\to \ell,\bm{b})}_{i\to j}},
    \end{align}
    where
    \begin{align}
        E^{(\ell,\bm{b})}_{ij} &= \abs{\tilde G^{(\bm{b},\ell)}_i \cap_m \tilde G^{(\bm{b},\ell)}_j} ,\\
        E^{(k\to \ell,\bm{b})}_{i\to j} &= \abs{\tilde G^{(\bm{b},k\to\ell)}_i \cap_m \tilde G^{(\bm{b},\ell)}_j} 
    \end{align}
    are the appropriately modified overlap measures, with $\tilde
    G^{(\bm{b},\ell)}_i$ the layer of hyperedges of size~$\ell$ in $\tilde
    G^{(\bm{b})}_i$, and $\tilde G^{(\bm{b},k\to \ell)}_i$ the projection
    of the layer $\tilde G^{(\bm{b},k)}_i$ onto hyperedges of size~$\ell$.
    We have also defined
    \begin{align}
        k_i^{(\ell,\bm{b})} = \argmin_{k\geq \ell}\left\{\log {E^{(k\to \ell)}_i \choose E^{(k\to \ell,\bm{b})}_{i\to j}}\multiset{\multiset{B}{\ell} }{ E^{(\ell)}_j-E^{(k\to \ell,\bm{b})}_{i\to j}} \right\}   
    \end{align}
    analogously to Eq.~(\ref{best-layer}). 

    \medskip

    It is worth noting that this mesoscale measure can be directly
    applied as a means to assess similarity among weighted
    hypergraphs, in the common case that the edges are positive
    integers (such as counts or frequencies in temporal systems). In
    this case, we can treat the hypergraphs as multi-hypergraphs with
    an edge $e$ of weight $w(e)$ corresponding to $w(e)$ independent
    hyperedges on the tuple of nodes $e$. We then set the partition
    $\bm{b}$ for each hypergraph to be the partition into $N$ groups
    of size $1$, such that the labels of the nodes are unchanged. This
    allows the encodings to properly account for the multi-edges when
    assessing similarity, by using the multiset combinatorics
    described above.

    \medskip

\begin{figure}[t]
    \includegraphics[width=0.70\textwidth]{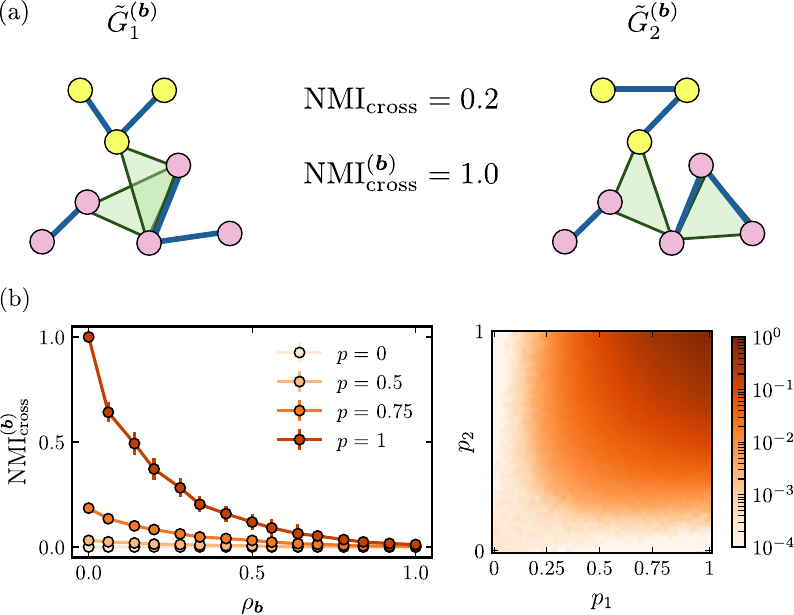}
    \caption{
        \textbf{Mesoscale similarity for hypergraphs.} 
        (a)~$\text{NMI}_{\rm cross}$ and its mesoscale variant
        $\text{NMI}^{(\bm{b})}_{\rm cross}$ for two small example networks
        on $N=8$ nodes, with the partition $\bm{b}$ dividing the nodes
        into $B=2$ groups indicated in yellow and pink. While the
        mesoscale measure is able to detect perfect similarity among the
        coarse-grained hypergraphs~$\tilde G_1^{({\bm b})}$ and $\tilde
        G_2^{({\bm b})}$, the standard NMI variant detects a low level of
        similarity at the node-level.
        (b) Mesoscale NMI for pairs of random clustered hypergraphs
        generated with an average fraction $p$ of nodes belonging to the
        same group. As we increase the level of noise $\rho_{\bm{b}}$
        between the two hypergraphs' underlying node partitions, the
        mesoscale NMI smoothly decreases, with stronger levels of
        community structure $p$ resulting in a more severe decline in the
        NMI (left). When both hypergraphs are generated from the same
        underlying node partition ($\rho_{\bm b}=0$) with different
        community strengths $p_1,p_2$, we see that greater levels of
        community structure result in greater levels of shared information
        among the hypergraphs, with $p_1=p_2=1$ giving maximum similarity. 
    }
    \label{fig:multiscale}
    \end{figure}

    In Fig.~\ref{fig:multiscale}(a) we show an illustration of the regular
    and mesoscale variants of $\text{NMI}_{\text{cross}}$ between two
    small example hypergraphs, with communities~$\bm{b}$ indicated in
    yellow and pink. When ignoring the node partition~$\bm{b}$, the NMI is
    quite low ($0.2$), as there is little structural overlap among the
    hypergraphs at the node-level. However, when we apply
    $\text{NMI}^{(\bm{b})}_{\text{cross}}$, we find maximum similarity due
    to identical coarse-grained representations $\tilde G^{(\bm{b})}_i$ at
    the mesoscale, which are not captured by the regular NMI measure.

    \medskip

    \medskip

    We then run simulations using synthetic hypergraph pairs $G_1,G_2$ on
    $N=1000$ nodes, tuning the level of planted community structure and
    level of similarity in their underlying node
    partitions~$\bm{b}^{(1)},\bm{b}^{(2)}$,
    which we set to have $B=50$ groups. We fix the layer sizes to
    $E^{(\ell)}=2^{12-\ell}$ for $\ell\in\{2,\dots,10\}$ and generate each
    $\ell$-hyperedge through repetition of the following process
    $E^{(\ell)}$ times: 

    \medskip

    \begin{enumerate}
        \item Choose a group $r\in \{1,\dots,B\}$ at random to form the
        majority affiliation of a new hyperedge. 
        \item Generate group affiliations for each of the
        remaining~$\ell-1$ nodes by picking group~$r$ with probability~$p$
        and another community label~$s\neq r$ uniformly at random from the
        remaining labels with probability~$1-p$.
        \item For each community label in the hyperedge, pick a node from
        that community uniformly at random, without replacement.
    \end{enumerate}

    \medskip

    This process results in random hypergraphs in which the expected
    fraction of nodes belonging to the majority community in each
    hyperedge is $p$. In this way, the individual hyperedges are
    independent and randomized across $G_1,G_2$, so we would expect NMI
    values near zero for the three original similarity measures discussed
    in Sec.~\ref{sec:hierarchy}. However, the mesoscale~NMI
    measure~$\text{NMI}^{(\bm{b}^{(1)})}_{\text{cross}}$ should be able to
    detect the similarity among the generated hypergraphs at the level of
    the planted modular structure. We expect that as the graphs $G_1,G_2$
    become less modular---i.e., the level of community strength~$p$
    decreases---the mesoscale similarity should decrease.  

    \medskip

    We can also vary the extent to which the modular structure overlaps
    across the two hypergraphs. For this, we take the partition~$b^{(1)}$
    used to generate $G_1$ and shuffle pairs of elements to form the
    partition~$\bm{b}^{(2)}$ which is used to generate $G_2$. We use a
    parameter~$\rho_{\bm{b}}$ to tune this shuffling, with
    $\rho_{\bm{b}}=0$ corresponding to no shuffling and $\rho_{\bm{b}}=1$
    correspond to swapping $N/2$ pairs of elements, so that all community
    labels have been perturbed. 

    \medskip

    In Fig.~\ref{fig:multiscale}(b) we show the results of these
    experiments. On the left we plot the mesoscale~NMI versus the level of
    partition noise~$\rho_{\bm{b}}$ for hypergraph pairs with various
    levels of community strength $p$. Markers are again averages over ten
    trials, with error bars representing three standard errors in the
    mean. We can see that the mesoscale NMI measure attributes maximum
    similarity for maximal community strength~$p=1$ when the partitions
    are not shuffled~($\rho_{\bm{b}}=0$). We can also see that it
    attributes a similarity of nearly zero for all $\rho_{\bm{b}}$ when
    there is very weak community structure~($p=0$). As we decrease the
    strength of community structure~$p$, we interpolate between these two
    regimes, with smooth decreases in similarity for greater levels of
    partition noise in all cases. On the right of
    Fig.~\ref{fig:multiscale}(b), we allow the community strengths~$p_1$
    and $p_2$ to be different between the two hypergraphs for
    $\rho_{\bm{b}}=0$, finding that mesoscale similarity is detected at
    high levels until around $p_i\approx 0.25$. This may be indicating a
    ``detectability transition''~\cite{DKMZ11a} in the planted community
    structure, in which the hyperedges are no longer correlated with the
    underlying shared node partition $\bm{b}$ in any meaningful way,
    resulting in a vanishing mesoscale~NMI. 

    \clearpage

    \section{Similarity of hypergraphs with tunable nestedness}\label{section:similaritynestedness}

\begin{figure}[h]
    \includegraphics[width=1.0\textwidth]{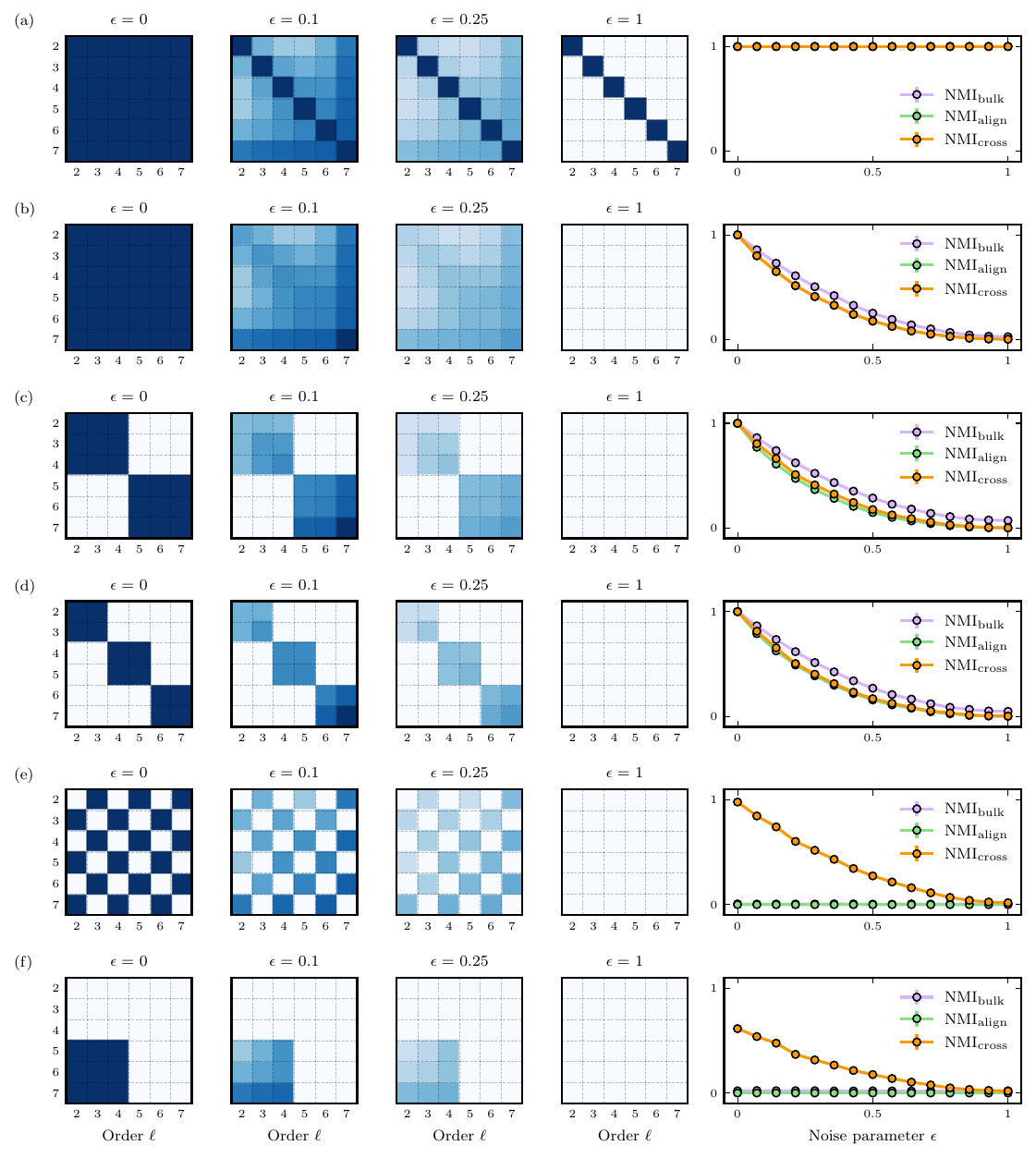}
    \caption{
        \textbf{Similarity scores against noise parameter for hypergraphs with tunable nestedness.}
        (a)~Fully nested hypergraphs dependently attacked.
        (b)~Fully nested hypergraphs independently attacked.
        (c)~2-block-nested hypergraphs.
        (d)~3-block-nested hypergraphs.
        (e)~Intertwined hypergraphs.
        (f)~Anti-block-nested hypergraph.
    }
    \label{fig:tunable}
    \end{figure}

\begin{figure*}[h!]
    \centering
    \includegraphics[width=1.00\textwidth]{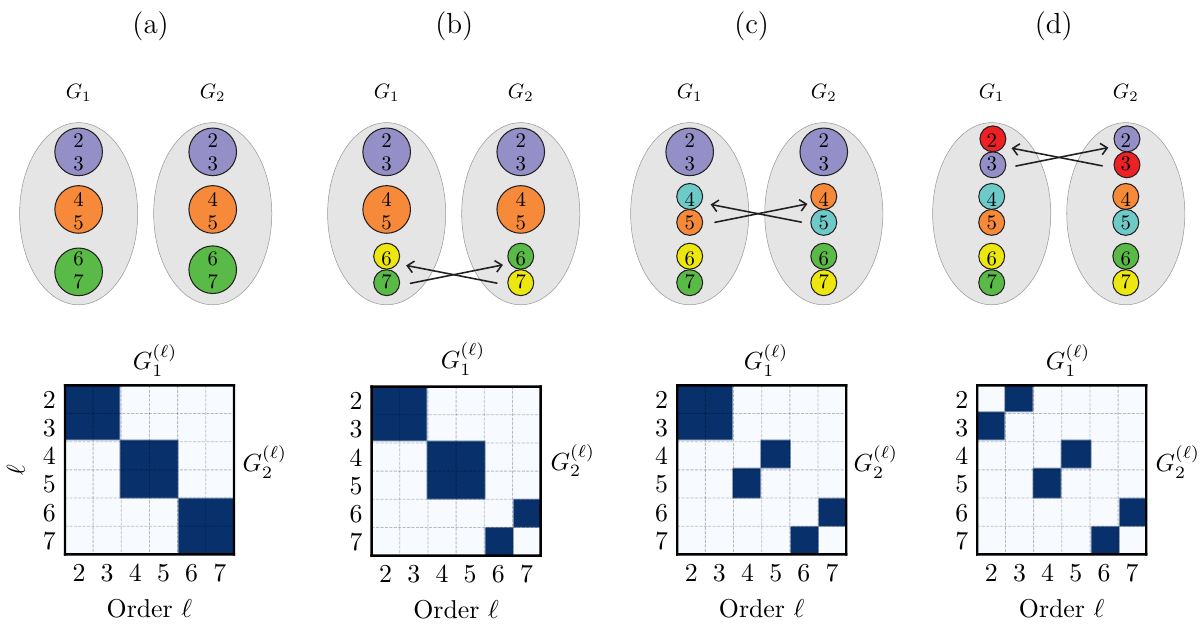}
    \caption{
    \textcolor{black}{
        \textbf{Randomization procedure of the block-nested hypergraph model.}
        (a)~We generate ``parent'' layers $\ell\in\{3,5,7\}$ in both $G_1$
        and $G_2$ and populate ``child'' layers $\ell'\in\{2,4,6\}$ as the
        projections of $G_i^{(\ell=\ell'+1)}$ in the \emph{other}
        hypergraph $G_j^{(\ell')}$.  That is, layers 2, 4, and 6 of $G_2$
        are nested in layers 3, 5, and 7 of $G_1$ (and vice-versa in this
        initial setting).
        (b)~We generate parent layer $\ell=7$ independently at random in
        $G_1,G_2$, then generate the child layers $\ell=6$ in $G_2,G_1$
        respectively as projections of these parent layers at $\ell=7$.
        Arrows point from parent layer to child layer.
        (c)~We start with the system in (b) and generate parent layer
        $\ell=5$ independently at random in $G_1,G_2$, then generate the
        child layers $\ell=4$ in $G_2,G_1$ respectively as projections of
        these parent layers at $\ell=5$.
        (d)~We start with the system in (c) and generate parent layer
        $\ell=3$ independently at random in $G_1,G_2$, generating child
        layers as before. Intra-order similarity between hypergraphs is
        destroyed and only cross-order similarity remains, as seen in the
        heatmaps of the lower row (reproduced from the main text).
        }
    }
    \label{fig:blocks}
    \end{figure*}

    Here, we extend the analysis of intra- and cross-order similarity for
    more complex models of synthetic hypergraphs with tunable levels of
    nestedness. In particular, we compute the ${\rm NMI}_{\rm bulk}$,
    ${\rm NMI}_{\rm align}$, and ${\rm NMI}_{\rm cross}$ of $N=100$ node
    hypergraphs under various levels of noise $\epsilon$.  We describe the
    models and experiments below, followed by Fig.~\ref{fig:tunable}
    showing the results.  At the end of this section
    (Fig.~\ref{fig:blocks}), we also illustrate in detail the
    randomization procedure for a paradigmatic example of synthetic
    hypergraph, i.e.  the block-nested hypergraph model used to highlight
    cross-order similarity in the absence of intra-order similarity.
    Similar procedures are employed to generate the other synthetic
    structures.

    \begin{itemize}
        \item Fully nested hypergraphs with identical noise.  We
        initialize two hypergraphs $G_1$ and $G_2$ over the same set of
        $N=100$ nodes.  We then generate, independently at random,
        interactions of order $\ell=7$.  Interactions of lower
        orders~$\ell\in\{2,3,4,5,6\}$ are generated by selecting all
        tuples of nodes which are subsets of the tuples encoding
        interactions of order~7.  The layers of interaction are assigned
        to $G_1$ and $G_2$, making them identical fully nested
        hypergraphs.  We add noise to both $G_1$ and $G_2$ at the same
        noise level $\epsilon$, rewiring all orders of interaction
        identically in each hypergraph. In this case, layers of the same
        order are kept identical across hypergraphs while layers of
        different orders become uncorrelated within the hypergraphs.  All
        three NMI scores indicate perfect similarity, as expected.

        \item Fully nested hypergraphs with independent noise.  Same as
        the previous model, but the layers of hypergraphs $G_1$ and $G_2$
        are independently rewired.  All NMI measures start at the maximum
        similarity, but smoothly decay to zero since layers of all sizes
        become uncorrelated across the hypergraphs.
       
        \item 2-block-nested hypergraphs.  We generate, independently at
        random, interactions of order $\ell=4$. Interactions of order 2
        and 3 are generated by selecting all tuples of nodes which are
        subsets of the tuples of order 4. Analogously, we generate
        independently at random interactions of order 7, and generate
        orders 5 and 6 by selecting tuples which are all subsets of
        interactions at layer~7.  The layers are then independently
        attacked such that their shared block-structure is destroyed. All
        NMI scores smoothly decrease with $\epsilon$.

        \item 3-block-nested hypergraphs.  Same procedure as previous
        model, but with a three-block architecture instead: layer 3
        generates layer 2; layer 5 generates layer 4; and layer 7
        generates layer 6.  Graphs are independently rewired and all
        scores smoothly decrease with $\epsilon$ as before.

        \item Intertwined hypergraphs.  We generate independently at
        random interactions of order 3, 5, and 7 in hypergraph $G_i$.  We
        then take the corresponding subsets of these layers and assign
        them, respectively, to layers 2, 4, and 6 of $G_j$.  We then
        attack both hypergraphs independently.  Since the hypergraphs
        never shared intra-order similarity, the ${\rm NMI}_{\rm bulk}$
        and ${\rm NMI}_{\rm align}$ assigns zero similarity throughout the
        whole noise process, whereas ${\rm NMI}_{\rm cross}$ is able to
        detect the shared structure embedded across different layers of
        the hypergraphs.

        \item Anti-block-nested hypergraphs.  We initialize two
        ``reference'' 2-block-nested hypergraphs $H_A$, $H_B$, where
        layers 4 and 7 generate, respectively, layers 2, 3, and 5, 6 in
        both $H_A$ and $H_B$.  We then assign the block layer
        $\ell\in\{2,3,4\}$ from $H_A$ to $G_1$, and $\ell\in\{5,6,7\}$
        from $H_B$ to $G_1$. Analogously, we take the block layer
        $\ell\in\{2,3,4\}$ from $H_B$ to $G_2$, and $\ell\in\{5,6,7\}$
        from $H_A$ to $G_2$.  Both graphs are independently attacked.
        Only the ${\rm NMI}_{\rm cross}$ measure is able to detect shared
        similarity prior to the full rewiring at $\epsilon=1$.
    \end{itemize}

    Throughout the experiments, the density of hyperedges is kept
    meaningful across all layers of interactions, in the sense that the
    size of layer $\ell$ is set at $E^{(\ell)}=E^{(\ell_{\rm
    max})}\binom{\ell_{\rm max}}{\ell}$ with a choice of $E^{(\ell_{\rm
    max})}\geq 100$.

    \clearpage

    \clearpage

    \clearpage

    \section{Empirical multiplex hypergraphs}\label{section:multiplex}

    Here we present summary statistics and pre-processing details for the
    three multiplex hypergraph datasets shown in the main text,
    representing scientific collaboration (APS physics
    fields~\cite{pacs_data}), movie co-appearances (IMDb movie
    genres~\cite{lotito2024multiplex}), and software development teams
    (Rust Github
    repositories~\cite{schueller2022evolving,schueller2024modeling,betti2025dynamics}).

    \medskip

    The APS multiplex dataset~\cite{pacs_data} contains ten layers
    representing ten physics fields according to the Physics and Astronomy
    Classification Scheme (PACS) of the American Physical Society (APS).
    Each layer-field is a hypergraph in which actors are nodes connected
    via hyperedges representing a paper published in that particular
    field.  For instance, a paper with three authors in Nuclear Physics is
    a hyperedge of size three in the corresponding layer ``NPhy''.  Layers
    vary in terms of number of nodes~$N$, total number of hyperedges
    edges~$E$, and maximum order of interaction~$\ell_{\rm max}$. For
    example, the condensed matter subfields (CM1 and CM2) tend to have
    papers with only a few authors, while the Elementary Particles (EPart)
    layer has some papers with thousands of authors. See
    Table~\ref{tab:table_aps} for further details.

    \medskip

    The IMDb multiplex dataset contains eight layers representing eight
    movie genres according to the Internet Movie Database (IMDd). Each
    layer is a hypergraph in which actors are nodes connected via
    hyperedges representing their co-appearance within a movie of the
    corresponding genre (see Table~\ref{tab:table_imdb}).

    \medskip

    The Github multiplex contains ten layers representing ten categories
    from the Rust Github repositories. Each layer-repository is a
    hypergraph in which users are nodes connected via hyperedges
    representing collaboration on a project in the corresponding category
    (see Table~\ref{tab:table_github}).

    \medskip

\begin{figure}[h]
    \includegraphics[width=0.85\textwidth]{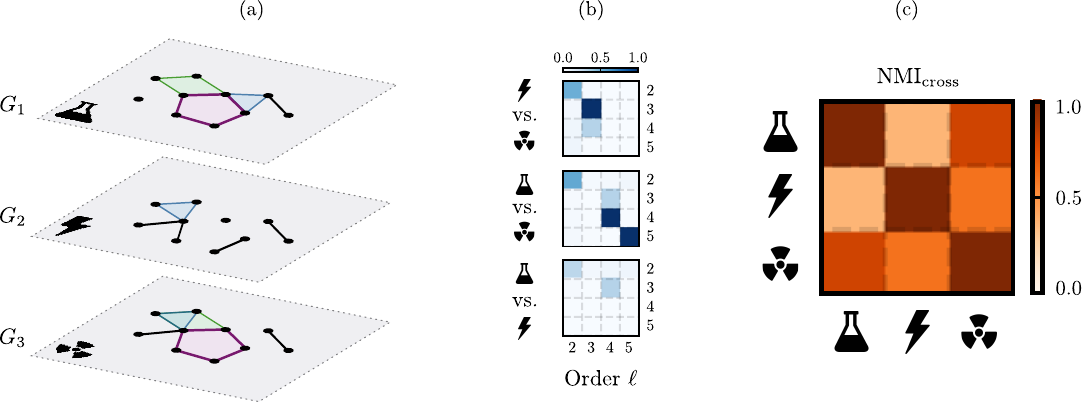}
    \caption{
        \textbf{Preprocessing and analysis of empirical datasets.}
        (a) Diagram of multiplex hypergraph containing three layers $G_1$, $G_2$, and $G_3$.
        (b)~Order-order similarity heatmaps of the multiplex layers.
        (c)~${\rm NMI}_{\rm cross}$ between layers of the multiplex.
    }
    \label{fig:figS1}
    \end{figure}

\begin{figure}[h!]
    \includegraphics[width=0.785\textwidth]{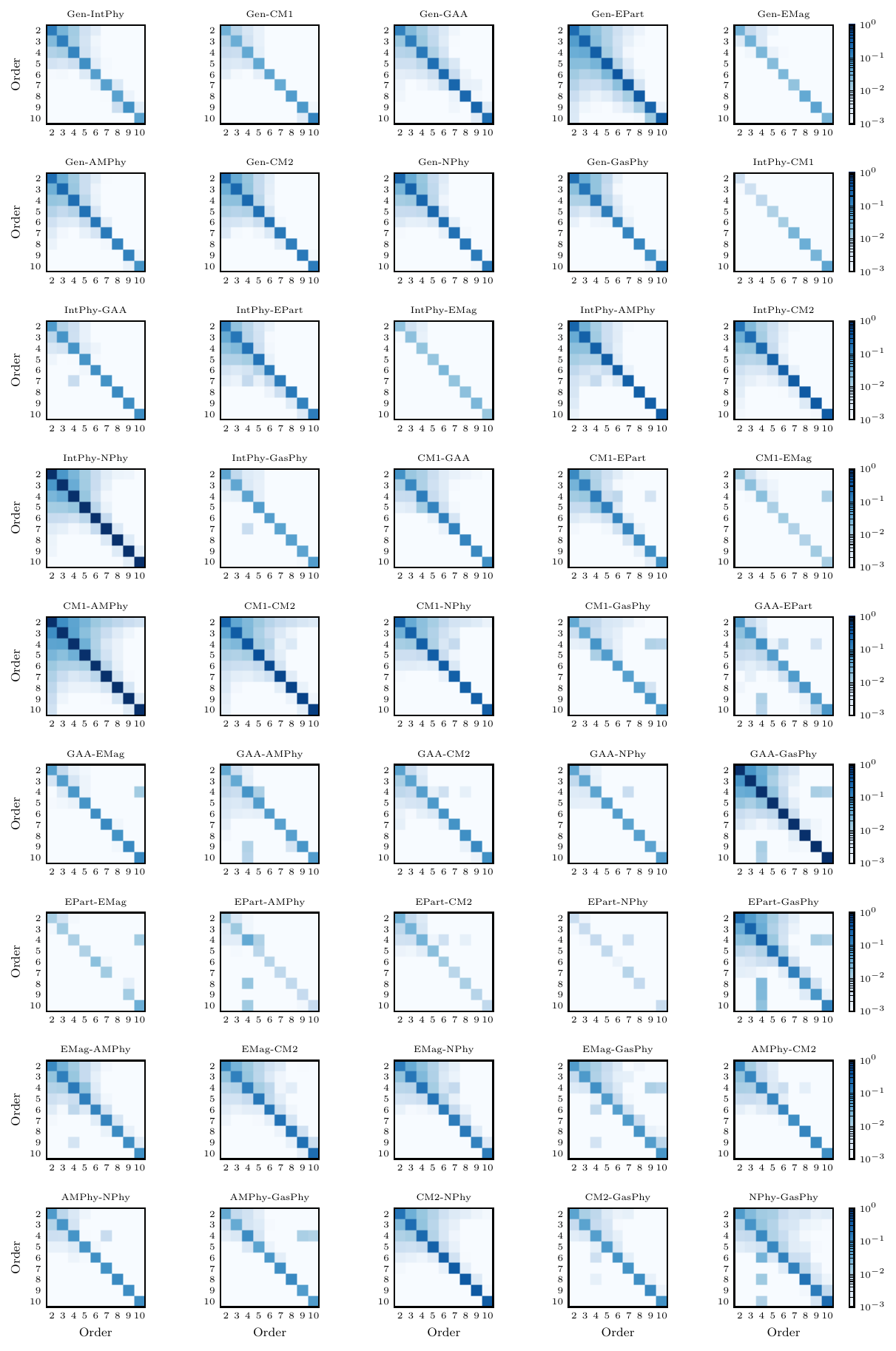}
    \caption{
        \textbf{Order-order similarity matrices of the APS physics fields dataset.}
    }
    \label{fig:figS_aps}
    \end{figure}

\begin{figure}[h!]
    \includegraphics[width=1.0\textwidth]{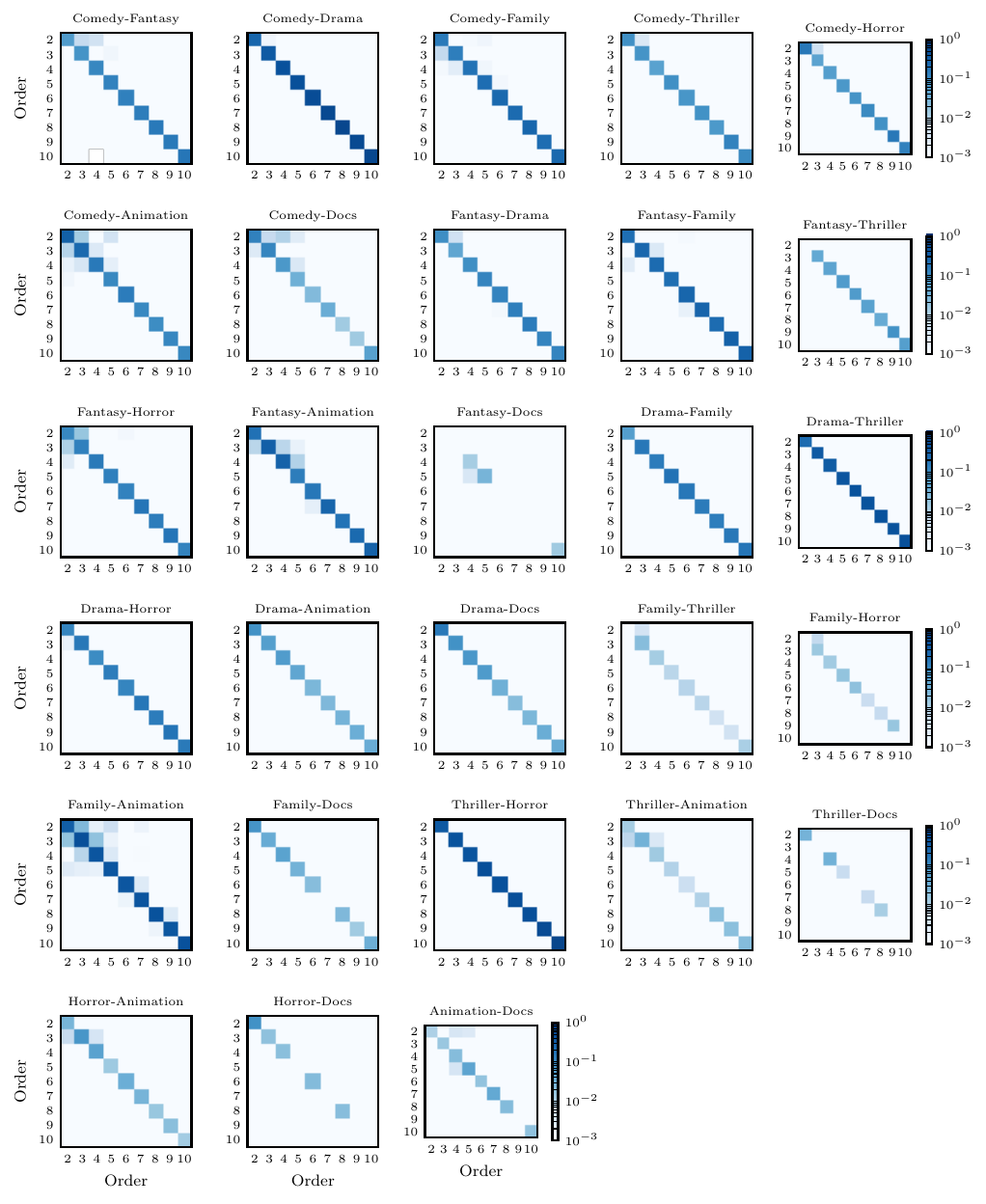}
    \caption{
        \textbf{Order-order similarity matrices of the IMBd movie genres dataset.}
    }
    \label{fig:figS_imdb}
    \end{figure}

\begin{figure}[h!]
    \includegraphics[width=0.825\textwidth]{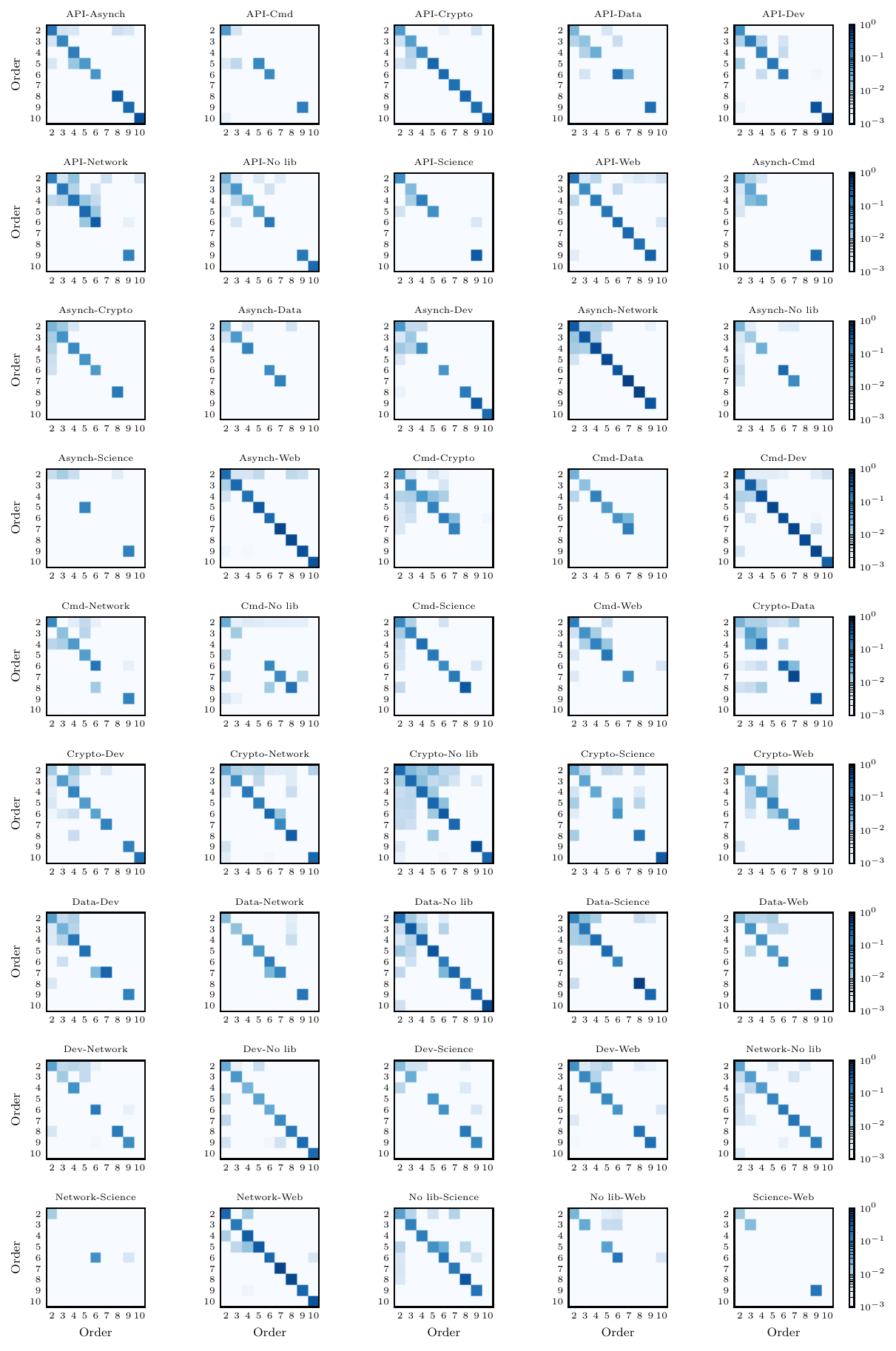}
    \caption{
        \textbf{Order-order similarity matrices of the Rust Github repository dataset.}
    }
    \label{fig:figS_git}
    \end{figure}

    For each dataset, we considered only nodes that co-appeared in at
    least two different layers, allowing for the presence of cross-order
    overlap. As described in the main text, we then computed the
    $\text{NMI}_{\rm cross}$ score between each layer of the multiplex in
    order to assess the similarity of physics fields, movie genres, and
    repository categories (Fig.~\ref{fig:empirical} in the main text).
    Figure~\ref{fig:figS1} illustrates our preprocessing and analysis of
    the empirical multiplex hypergraphs.  Below we show the results of
    computing the pairwise similarity for the different orders of
    interaction $\ell=2,\dots,10$ in the same manner as in
    Figs.~\ref{fig:fig1}-\ref{fig:bulkalign}.  Figure~\ref{fig:figS_aps}
    shows the similarity between orders of interaction for all
    combinations of physics fields.  Most PAC pairs show high similarity
    scores only for lower-order interactions, with the exception of a few
    pairs such as the condensed matter fields and nuclear and
    interdisciplinary physics (NPhy and IntPhy).  Similar results are
    shown for the IMDb and Github datasets in Figs.~\ref{fig:figS_imdb}
    and \ref{fig:figS_git}, respectively.

    \medskip

    Finally, we highlight that throughout our empirical analysis we took
    the intersection of the node sets in $G_1$ and $G_2$ as their common
    node set of size $N$. This allowed for comparisons only with respect
    to the nodes that actively participate in both hypergraphs, which is
    preferable if some nodes are naturally constrained to only exist in
    one of the two hypergraphs. For example, in the scientific
    co-authorship hypergraphs, we focus on interdisciplinary authors that
    publish papers in multiple disciplines (each discipline being an
    independent hypergraph). In this case, since a large portion of
    authors have short academic careers confined to a single discipline,
    and the frequency of attrition is discipline-dependent, it is more
    sensible to compare hypergraphs based on the authors that are active
    in multiple disciplines to understand structural similarities in
    collaboration patterns. An alternative option for hypergraphs of
    non-identical node sets is to use the union of the node sets in $G_1$
    and $G_2$ as the shared node set of size $N$, which requires adding
    isolated nodes to one or both node sets until they match. This
    approach is preferable when comparing systems in which the absence of
    nodes provides important evidence of structural dissimilarity, since
    it provides an increasingly strong penalty on the NMI as the node sets
    of $G_1,G_2$ overlap less. Either choice of preprocessing is
    compatible with the specific encodings we present in
    Sec.~\ref{sec:hierarchy}.

    \clearpage

    \begin{table}[h!]
        \centering
        \renewcommand{\arraystretch}{1.3} 
        \begin{tabular}{l@{\hspace{40pt}}ccc @{\hspace{15pt}} cc} 
        \toprule 
        \textbf{PACS} & \multicolumn{1}{c}{$N$} & \multicolumn{1}{c}{$E$} & $\ell_{\rm max}$ & $N^{(\ell \leq 10)}$ & $E^{(\ell \leq 10)}$ \\
        \midrule 
        General (Gen) & 87712 & 48751 & 2926 & 74360 & 48077 \\
        Elementary Particles (EPart) & 49920 & 18913 & 3047 & 19550 & 15256 \\
        Nuclear Physics (NPhy) & 41335 & 14407 & 2895 & 14444 & 10285 \\
        Atomic and Molecular Physics (AMPhy) & 36414 & 15599 & 76 & 32258 & 14892 \\
        Electromagnetism (EMag) & 62335 & 37447 & 146 & 57502 & 36238 \\
        Physics of Gases (GasPhy) & 12693 & 4031 & 409 & 9182 & 3554 \\
        Condensed Matter: Thermal Properties (CM1) & 83351 & 40315 & 65 & 77596 & 38913 \\
        Condensed Matter: Optical Properties (CM2) & 91267 & 51890 & 131 & 84001 & 49063 \\
        Interdisciplinary Physics (IntPhy) & 71801 & 28823 & 627 & 65663 & 28065 \\
        Geophysics, Astronomy, and Astrophysics (GAA) & 55975 & 14393 & 2921 & 29903 & 13542 \\
        \bottomrule
        \end{tabular}
        \caption{\textbf{Statistics of the APS multiplex hypergraph.}}
        \label{tab:table_aps}
    \end{table}


    \begin{table}[h!]
        \centering
        \renewcommand{\arraystretch}{1.3} 
        \begin{tabular}{l@{\hspace{40pt}}ccc @{\hspace{15pt}} cc} 
        \toprule 
        \textbf{Genre} & \multicolumn{1}{c}{$N$} & \multicolumn{1}{c}{$E$} & $\ell_{\rm max}$ & \multicolumn{1}{c}{$N^{(\ell \leq 10)}$} & \multicolumn{1}{c}{$E^{(\ell \leq 10)}$} \\
        \midrule 
        Comedy & 58432 & 7992 & 313 & 15317 & 3190 \\
        Animation & 8322 & 1372 & 100 & 3103 & 741 \\
        Family & 21787 & 2465 & 313 & 6028 & 1168 \\
        Fantasy & 21366 & 1925 & 313 & 5020 & 818 \\
        Drama & 71894 & 10957 & 224 & 20755 & 4615 \\
        Thriller & 49407 & 5905 & 158 & 12865 & 2369 \\
        Horror & 30084 & 3173 & 95 & 8422 & 1359 \\
        Documentary & 3570 & 482 & 112 & 1316 & 286 \\
        \bottomrule
        \end{tabular}
        \caption{\textbf{Statistics of the IMDb multiplex hypergraph.}}
        \label{tab:table_imdb}
    \end{table}

    \begin{table}[h!]
        \centering
        \renewcommand{\arraystretch}{1.3} 
        \begin{tabular}{l@{\hspace{40pt}}ccc @{\hspace{15pt}} cc} 
        \toprule 
        \textbf{Repository} & \multicolumn{1}{c}{$N$} & \multicolumn{1}{c}{$E$} & $\ell_{\rm max}$ & \multicolumn{1}{c}{$N^{(\ell \leq 10)}$} & \multicolumn{1}{c}{$E^{(\ell \leq 10)}$} \\
        \midrule 
        API bindings (API) & 1384 & 276 & 197 & 669 & 243 \\
        Asynchronous (Asynch) & 964 & 35 & 50 & 612 & 216 \\
        Command line utilities (Cmd) & 1216 & 275 & 113 & 719 & 253 \\
        Cryptography (Crypto) & 925 & 25 & 76 & 460 & 205 \\
        Data structures (Data) & 889 & 22 & 92 & 504 & 200 \\
        Development tools (Dev) & 1594 & 370 & 76 & 846 & 326 \\
        Network programming (Network) & 1090 & 266 & 64 & 714 & 250 \\
        No standard library (No lib) & 1255 & 381 & 92 & 627 & 329 \\
        Science & 701 & 41 & 197 & 364 & 133 \\
        Web programming (Web) & 1003 & 236 & 64 & 620 & 219 \\
        \bottomrule
        \end{tabular}
        \caption{\textbf{Statistics of the Rust GitHub multiplex hypergraph.}}
        \label{tab:table_github}
    \end{table}

    \clearpage

    \section{Similarity of empirical multiplex hypergraphs}\label{section:multiscale}

    \clearpage

    \clearpage

    \clearpage
    \section{Runtime scaling on empirical multiplex datasets}\label{section:runtime}

    We compute the average runtime required, per pair of hypergraphs
    $\{G_i,G_j\}$, to compute the similarity values
    $\text{NMI}_{\text{cross}}(G_i,G_j)$ used in the experiments of
    Fig.~\ref{fig:empirical}. For each multiplex dataset, we examine how
    this runtime scales with the maximum layer order $\ell_{\text{max}}$
    included for the analysis, which gives estimates of the empirical
    runtime scaling behavior of our measure.

    \medskip

\begin{figure}[h]
    \includegraphics[width=1\textwidth]{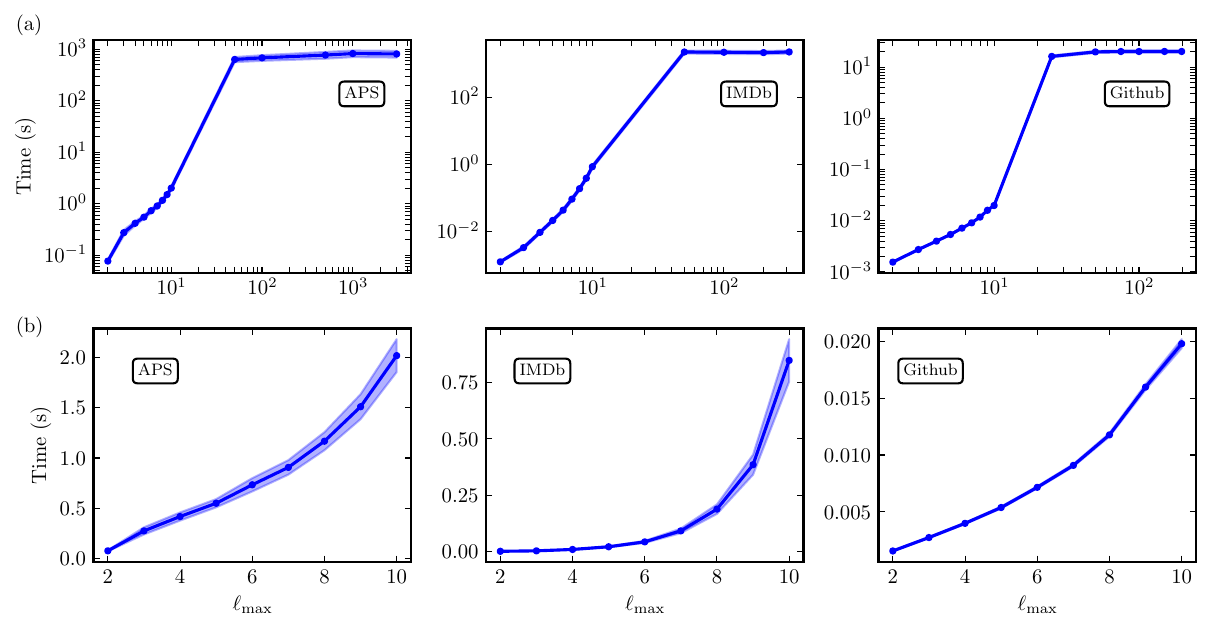}
    \caption{\textbf{Runtime scaling in real multiplex data. }}
    \label{fig:runtime}
    \end{figure}

    In Fig.~\ref{fig:runtime} we show the results of these experiments for
    the three multiplex datasets. In row (a) we plot the results for all
    layers, while in row (b) we zoom in on the range $\ell_{\text{max}}\in
    [2,10]$. We find that, as expected, the runtime scaling is roughly
    quadratic in $\ell_{\text{max}}$ for smaller values, in which all
    layers are occupied by hyperedges in most networks. We see slight
    deviations due to the number of edges in each layer---the hypothetical
    $O(\ell_{\text{max}}^2)$ scaling of of SI Section~\ref{app:algorithm}
    will only occur when all layers have an identical number of
    hyperedges. However, for very large maximum order $\ell_{\text{max}}$,
    we find that the runtime starts to level off. This is because the
    layers are much more sparsely occupied---in many cases, empty---for
    higher~$\ell$.

    \medskip

    Notably, the runtimes of $\text{NMI}_{\text{bulk}}$ and
    $\text{NMI}_{\text{align}}$ are negligible on all the empirical
    hypergraphs, due to not considering cross-layer contributions which
    require either explicit projection or recursive counting of nested
    overlaps. These experiments give a more realistic idea of how the
    proposed measures scale with the size of empirical hypergraphs,
    complementing the theoretical scaling results of SI
    Section~\ref{app:algorithm}. 

    \medskip

    \clearpage
    \section{Detecting anomalies in temporal hypergraph streams}\label{section:temporal}

    In longitudinal studies of social behavior in humans and other
    animals, one often looks for changepoints, anomalies, or other notable
    features in the temporal interaction dynamics, which are intrinsically
    higher-order in nature~\cite{cencetti2021temporal,iacopini2024not}.
    Applying the hypergraph similarity framework presented here to the
    temporal hypergraph snapshots of these systems allows for the
    detection of meaningful structural variability across time in such
    applications, which is not possible using standard graph similarity
    measures when groups vary in size over
    time~\cite{iacopini2024temporal}.

    \medskip

    Here we explore an application of our method for detecting anomalies
    in the Enron email dataset~\cite{klimt2004enron}, which is naturally
    represented as a temporal hypergraph in which nodes are email
    addresses and each hyperedge represents the sender and all receivers
    of a particular email~\cite{benson2018simplicial}. Hyperedges are
    timestamped according to the time the email was sent, allowing for the
    construction of hypergraph snapshots for different time periods. For
    these analyses, we bin the hyperedges into thirty-day periods to
    capture month-month fluctuations in email activity, but the general
    conclusions we find persist under different binnings. The emails took
    place over a period of roughly 45 months in the late 1990s to early
    2000s, during which the Enron corporation was involved in one of the
    largest accounting scandals in history. A number of works have aimed
    to understand the structure of these emails from the perspective of
    pairwise graphs~\cite{diesner2005communication,hardin2015network} and
    hypergraphs~\cite{benson2018simplicial}. 

    \medskip

\begin{figure}[htb!]
    \includegraphics[width=0.95\textwidth]{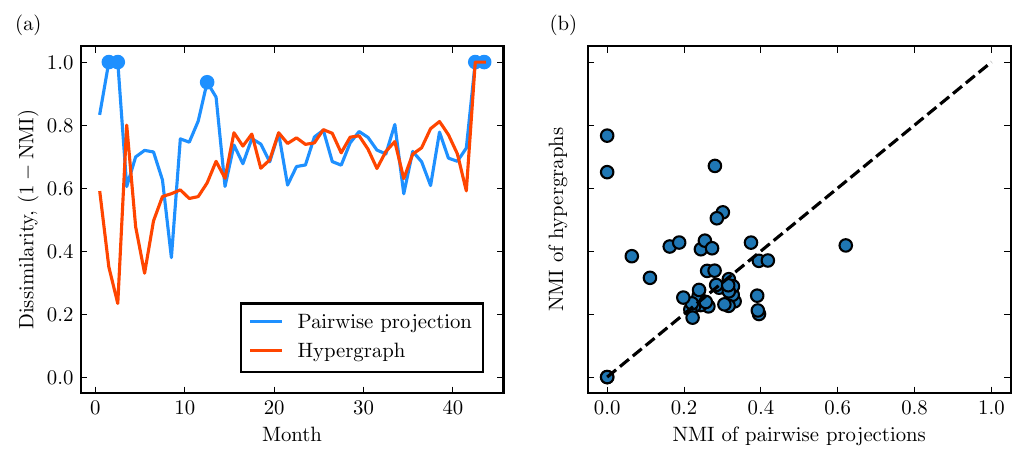}
    \caption{
        \textbf{Detecting anomalies in temporal graph data.} 
        (a)~Month-to-month dissimilarity $1-\text{NMI}(G_t,G_{t+1})$
        computed using the graph similarity~\cite{felippe2024network}
        among the pairwise projections (blue) as well as the hypergraph
        similarity ($\text{NMI}_{\text{cross}}$, red), for the Enron email
        dataset~\cite{benson2018simplicial}.
        (b)~Scatterplot of the time series values produced by both
        methods, with Pearson and Spearman correlation values of $-0.04$
        and $0.01$ respectively.
        }
    \label{fig:enron}
    \end{figure}

    In Fig.~\ref{fig:enron}(a) we show the
    dissimilarity~$1-\text{NMI}(G_t,G_{t+1})$ among the emails from month
    $t$ to month $t+1$, for all months $t$ in the dataset. One curve shows
    the results obtained by computing the NMI using the pairwise
    projection of the hypergraph at time $t$ and the graph NMI measure
    of~\cite{felippe2024network}, while the other curve shows the result
    of computing the NMI using the $\text{NMI}_{\text{cross}}$ measure we
    propose here. We also identify outliers in each time series using the
    crude (but widely used) interquartile range (IQR) method, in which any
    data point that exceeds the third quartile by more than $1.5$ IQRs is
    considered a high outlier. Such high outliers in this case---i.e.,
    anomalously high dissimilarity values---may correspond to abrupt
    shifts in the network structure of the emails, signifying an
    organizational change. These anomalies are highlighted as circular
    markers. 

    \medskip

    We can see that the time series constructed using the pairwise and
    hypergraph similarity measures share some underlying fluctuations but
    are largely uncorrelated, with only the pairwise series having
    anomalies according to the IQR method. This is because the pairwise
    measure does not capture the nested structure of the interactions,
    causing it to underestimate similarity in instances where hyperedges
    merge and split up over time. We also find qualitatively different
    autocorrelation structure among the two series: the hypergraph NMI
    time series has moderate to high positive autocorrelation for lags up
    to five months
    ($\{\rho(1),\rho(2),\rho(3),\rho(4),\rho(5)\}=\{0.49,0.24,0.40,0.44,0.35\}$),
    while the pairwise NMI time series only has positive autocorrelation
        for a lag of one month
            ($\{\rho(1),\rho(2),\rho(3),\rho(4),\rho(5)\}=\{0.31,-0.03,-0.09,-0.12,-0.09\}$).
            In Fig.~\ref{fig:enron}(b) we plot the time series values as a
            scatterplot, which shows the lack of correlation among the
            series constructed using the pairwise and hypergraph NMI
            measures. The Pearson and Spearman correlation coefficients
            between the two series are $-0.03$ and $0.01$ respectively. 

    \medskip

    These results provide an example of how, by enabling the detection of
    more nuanced aspects of similarity among datasets consisting of
    higher-order interactions, the proposed hypergraph similarity
    framework can provide qualitatively different conclusions in
    real-world application scenarios.

    \clearpage
    \textcolor{black}{\section{Hypergraph similarity and contagion dynamics}\label{section:dynamcis}}

    Here we investigate quantitatively how structural similarity between
    two hypergraphs, as measured by our $\text{NMI}_{\rm cross}$ measure,
    is linked to the outcome of a dynamical process running on the two
    systems. To perform such an analysis in a systematic way, we construct
    a fully nested hypergraph $G$ --- which will be used as our reference
    hypergraph --- and progressively perturb its higher–order structure to
    create a second hypergraph $G'$. The perturbation consists of
    randomizing interactions independently across orders with probability
    $\epsilon \in [0,1]$, where $\epsilon=0$ corresponds to the original
    nested configurati and $\epsilon=1$ corresponds to a fully randomized
    structure. Importantly, during this process we preserve the degree and
    hyper-degree distributions so that only the organization of
    interactions changes while the local connectivity statistics remain
    fixed. We consider regular hypergraphs with $N=900$ nodes and a
    maximum number of layers $\ell_{\rm max}=3$, the number of pairwise
    interactions and triplets fixed to $E^{(2)}=4050$ and $E^{(3)}=600$,
    respectively. The degree and hyper-degree distributions are also fixed
    at $k_1=9$ and $k_2=2$, such that these distributions remain unchanged
    throughout the entire perturbation procedure.

    \medskip

\begin{figure*}[h!]
    \centering
    \includegraphics[width=1.00\textwidth]{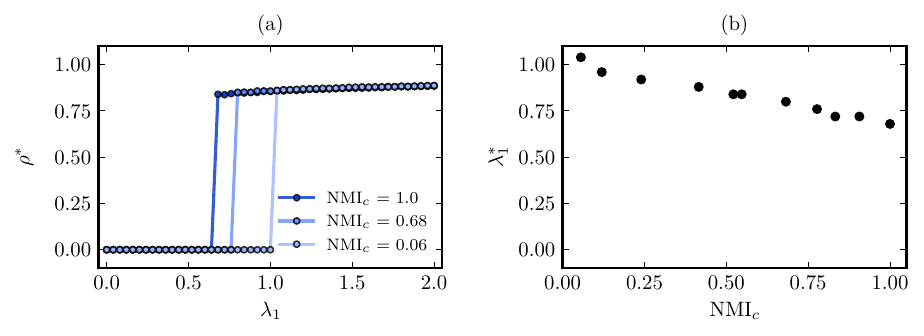}
    \caption{
        \textbf{Epidemic onset and hypergraph similarity.} 
        (a)~Stationary prevalence $\rho^*$ versus infectivity $\lambda_1$
        for different structural perturbations of a nested hypergraph $G$.
        The cross-order NMI between $G$ and its perturbed variant
        $G'$ for different levels of perturbation noise are shown.
        (b)~Epidemic threshold $\lambda_1^*$ for the perturbed graph $G'$
        as a function of its NMI with $G$.
    }
    \label{fig:figSIS}
    \end{figure*}

    We then simulate a higher–order SIS contagion
    process~\cite{iacopini2019simplicial} on the hypergraphs $G,G'$ and
    study how the stationary prevalence state $\rho^*$ varies as a
    function of the infectivity parameter $\lambda_1$. For each level of
    structural perturbation $\epsilon$ we estimate the epidemic threshold
    $\lambda_1^*$ and compute the hyperaph similarity
    $\text{NMI}_{\text{cross}}(G,G')$. We can observe in
    Fig.~\ref{fig:figSIS}(a) that the onset of the epidemic depends
    strongly on the structural organization of interactions: hypergraphs
    that are closer to the original nested configuration of $G$ exhibit an
    earlier transition, while increasingly randomized structures delay the
    epidemic onset.

    \medskip

    In Fig.~\ref{fig:figSIS}(b) we plot the critical value $\lambda_1^*$
    as a function of similarity with the initial nested hypergraph. The
    results show a clear decreasing trend, indicating that as similarity
    to the original nested structure decreases, the epidemic threshold
    systematically shifts to larger values signaling a slower onset of
    contagion.  This provides direct evidence that the proposed similarity
    measure captures structural features that are dynamically relevant for
    processes unfolding on higher–order networks.  Our findings are in
    agreement with previous work which has shown that continuously
    perturbing structuraligher-order features can affect the onset of
    collective behavior in higher–order dynamical processes
    models~\cite{malizia2025hyperedge}. 

    \clearpage
    \textcolor{black}{\section{Comparison with baseline and existing measures}\label{section:comparison}}

\begin{figure*}[h!]
    \centering
    \includegraphics[width=1.00\textwidth]{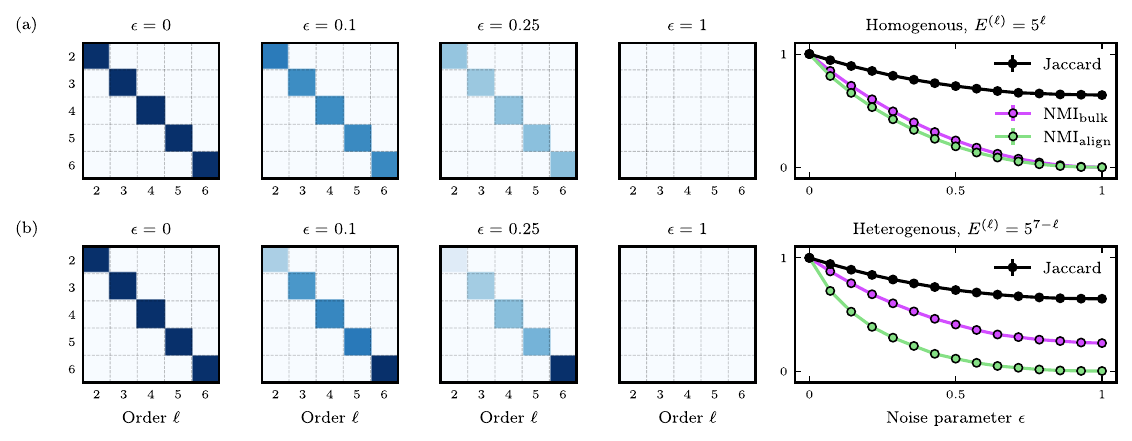}
    \caption{
        \textbf{Comparison of NMI measures (bulk and align) against the Jaccard similarity index averaged over the best hyperedge matching. }
        The Jaccard similarity severely inflates the similarity value for
        uncorrelated hypergraphs, due to spurious hyperedge overlaps
        attributable to the hyperedge densities.
        }
    \label{fig:nmi_vs_jaccard}
    \end{figure*}

    Here we compare our measures with a simple baseline and other recently
    proposed hypergraph similarity measures using the same synthetic tests
    as in the main text. A natural baseline for comparison is to take the
    average Jaccard similarity for the best hyperedge matching between the
    two hypergraphs  $G_1,G_2$, thus 
    \begin{align}
        s(G_1, G_2) = \frac{1}{2|G_1|} \sum_{e\in G_1} \textrm{max}_{e'\in G_2}
                                       \left\{ \frac{| e\cap e' |}{| e\cup e'|} \right\}
                      +
                      \frac{1}{2|G_2|} \sum_{e\in G_2} \textrm{max}_{e'\in G_1}
                                       \left\{ \frac{| e\cap e' |}{| e\cup e'|} \right\} ~ .
    \end{align}
    This measure satisfies $s(G_1,G_2)=1$ if and only if $G_1=G_2$, and
    will decrease as the hyperedge sets become more dissimilar in
    hyperedge-level overlap, having equal contributions from each
    hypergraph to the similarity score. In Fig.~\ref{fig:nmi_vs_jaccard},
    we reproduce Fig.~2 of the main text using this measure, finding that
    when compared to $\text{NMI}_{\rm bulk}$ and $\text{NMI}_{\rm align}$,
    this average Jaccard similarity measure has a similar smooth decrease
    with the noise level. However, it severely inflates the similarity of
    uncorrelated hypergraphs ($\epsilon=1$) which have structural overlap
    purely due to chance. The NMI measures naturally correct for this
    since structural overlaps that do not greatly exceed those expected by
    chance will fail to provide any compression when considering the
    shared information among the hyperedge sets. As discussed in the main
    text, the NMI-align measure is more effective than NMI-bulk for
    accounting for this baseline level of overlap in the presence of
    heterogeneous layer densities, hence the gap in the curves in panel
    (b).

\begin{figure*}[ht!]
    \centering
    \includegraphics[width=1.00\textwidth]{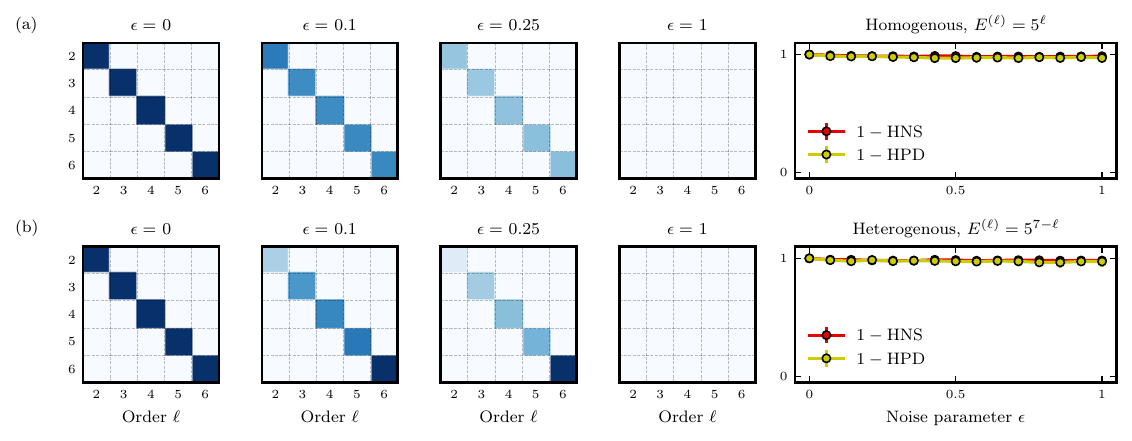}
    \caption{
        \textbf{Synthetic experimental results for the Hyper NetSimile (HNS) and Hyperedge Portrait Divergence (HPD) distance measures, transformed as $1-\text{distance}$ to form similarity measures.}}
    \label{fig:hns-and-hpd}
    \end{figure*}

    We also perform the same experiments with the two hypergraph
    dissimilarity measures proposed in~\cite{agostinelli2025higher}.
    The first measure, Hyper NetSimile (HNS), is a generalization of the
    NetSimile measure \cite{berlingerio2012netsimile} wherein graph
    similarity is regarded as the distance between signature vectors
    associated to multiple network descriptors. After considering
    generalized features such as hyper-degree, hyper-clustering
    coefficient, and so on, the HNS is defined as the Canberra distance
    between signature vectors $\bm{v}_1$ and $\bm{v}_2$ of
    hypergraphs $G_1$ and $G_2$, respectively,
    \begin{align}
        \text{HNS}(G_1,G_2) = d_{\rm Canberra}(\bm{v}_1, \bm{v}_2) 
                            = \frac{1}{V}\sum_{j=1}^V \frac{|v_1^j - v_2^j|}{|v_1^j| + |v_2^j|} ~ , 
    \end{align}
    where the distance is normalized by $V=|\bm{v}_i|$.  The second
    dissimilarity measure, Hyperedge Portrait Divergence (HPD), is the
    Jensen-Shannon divergence between distributions $P(m,n,l,k)$
    associated to the number of hyperedges of size $m$ having $k$
    hyperedges of size $n$ at a distance $l$, such that $l=1$ if at least
    one node is shared by the two hyperedges.  In
    Fig.~\ref{fig:hns-and-hpd} we show the results of the
    \emph{complement} of HNS and HPD (that is, $1-\text{HNS}$ and
    $1-\text{HPD}$, resp.) for the same experiment of Fig.~2 of the main
    text. Notably, for all noise parameters $\epsilon$ the dissimilarity
    measures are unable to distinguish the two random hypergraphs,
    regardless of their hyperedge layer density. This is because
    similarity with these two measures is assessed at a global level based
    on structural statistics rather than at a local level based on node
    IDs, which has the benefit of not requiring node alignment but is
    unable to distinguish the actual node sets that form the hyperedges so
    is not suited for analyzing node-aligned systems. In this experiment,
    as noise is added, the hypergraphs continue to have similar structural
    statistics despite the decreasing overlap in the node identities
    within their hyperedges. This results in the persistent high
    similarity values we see in these two measures.

\begin{figure*}[h!]
    \centering
    \includegraphics[width=0.90\textwidth]{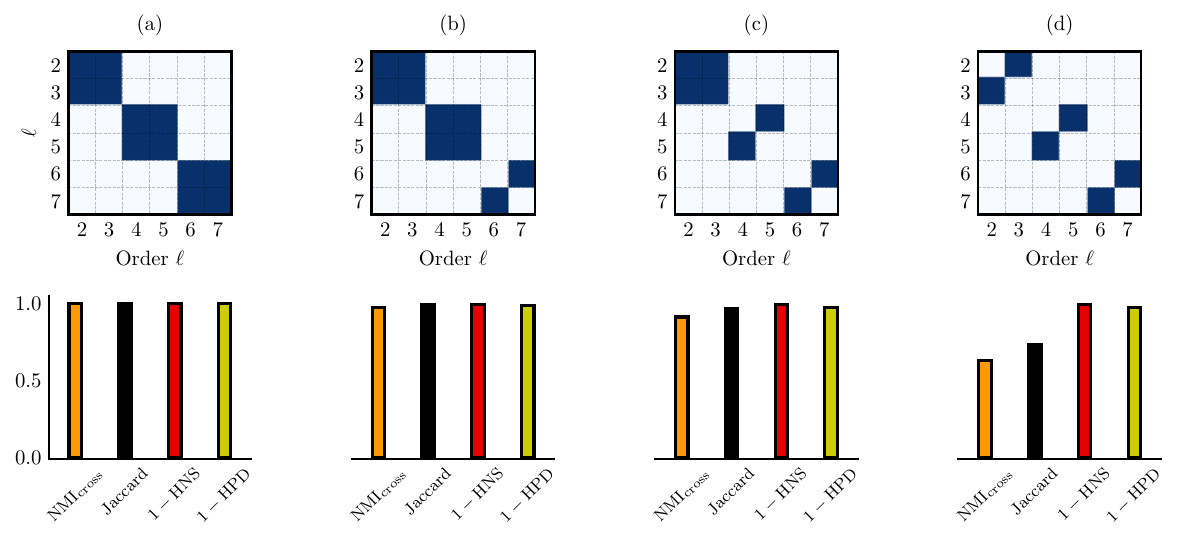}
    \caption{
        \textbf{Comparison of $\text{NMI}_{\rm cross}$ with the Jaccard baseline and measures of \cite{agostinelli2025higher} in the block-randomized hypergraph experiment of Fig.~3 in the main text.}
    }
    \label{fig:cross_comparison}
    \end{figure*}

    We also repeat the experiments of Fig.~3 of the main text, in which
    block-nested hypergraphs are sequentially attacked so as to highlight
    similarity across orders of interaction. For this, we compare our
    measure of $\text{NMI}_{\rm cross}$ against the three measures above
    for different block configurations (see
    Fig.~\ref{fig:cross_comparison}), finding a similar story. The
    average Jaccard similarity measure performs relatively well
    throughout the block disruptions, but still assigns more
    similarity than warranted. Meanwhile, both $1-\text{HNS}$ and
    $1-\text{HPD}$ are once again insensitive to any dissimilarity
    induced by the attacks over blocks.

\end{document}